\documentclass[fleqn,usenatbib]{mnras}
\usepackage{newtxtext,newtxmath}
\usepackage[T1]{fontenc}

\usepackage{multirow}
\usepackage{makecell} 
\usepackage{subcaption}
\usepackage{ulem}
\usepackage{soul}
\usepackage{graphicx}	%
\usepackage{amsmath}	
\usepackage{subcaption} 

\newcommand{\msun}{\,M$_{\odot}$}

\title[Weak shocks in the cool cores of clusters]{{X-ray shocks in the cool cores of galaxy clusters: insights from TNG-Cluster}}

\author[M. Prunier et al.]{
Marine Prunier$^{1,2,3}$\thanks{E-mail: prunier@mpia.de}
Francesco Ubertosi$^{4,5}$, Julie Hlavacek-Larrondo$^{2,3}$ and Annalisa Pillepich$^{1}$
\\
$^{1}$Max-Planck-Institut f{\"u}r Astronomie, K{\"o}nigstuhl 17, D-69117 Heidelberg, Germany\\
$^{2}$Département de Physique, Université de Montréal, Succ. Centre-Ville, Montréal, Québec, H3C 3J7, Canada\\
$^{3}$Centre de recherche en astrophysique du Quebec (CRAQ)\\
$^{4}$ Dipartimento di Fisica e Astronomia, Università di Bologna, via Gobetti 93/2, I-40129 Bologna, Italy\\
$^{5}$ Istituto Nazionale di Astrofisica - Istituto di Radioastronomia (IRA), via Gobetti 101, I-40129 Bologna, Italy}

\date{}
\pubyear{\the\year{}}

\begin{document}
\label{firstpage}
\maketitle

\begin{abstract} 
Shock fronts driven by active galactic nuclei in galaxy cluster cores represent a promising mechanism to heat the intracluster gas by converting kinetic energy into thermal energy through gas compression, thereby offsetting radiative cooling. Despite their potential importance, such shocks are challenging to detect, requiring deep X-ray exposures, and have only been identified in {ten massive clusters}. We present the first systematic detection and characterization of AGN-driven shocks in simulated clusters from the TNG-Cluster magnetohydrodynamic cosmological zoom-in simulations of galaxies. TNG-Cluster exhibits a rich variety of spatially-resolved X-ray structures, including realistic populations of X-ray cavities, as well as shocks, {produced by its AGN feedback model, without collimated, relativistic jets, nor cosmic rays.} We produce mock \textit{Chandra} observations with deep, 600-ks exposure times, for a selected sample of 100 clusters, mass-matched (M$_\text{500c}=1.2$ - $8.5 \times 10^{14}$ M$_\odot$) to the ten observed clusters with shocks. Using observational techniques, we identify shocks through surface brightness edges fitted with broken power laws and associated density and temperature jumps. We detect 50 shocks in 30 of the 100 clusters, with $\sim$35 per cent hosting multiple shocks. These shocks typically lie within a hundred kiloparsec of the central SMBH, are weak (Mach number $<$ 2, median $\sim$ 1.1), and {are associated with X-ray cavities in about half of the cases.} Both in observations and in TNG-Cluster, shocks tend to be located at larger radii than cavities, with median offsets of 46 kpc and 27 kpc, respectively. The observationally inferred shock powers are comparable to the cluster cooling luminosities (10$^{44-46}$ erg s$^{-1}$), suggesting that shocks in the simulation are a crucial heating mechanism. Our results {indicate} that shocks play a role as important as cavities in balancing cooling in cluster cores, acting isotropically and up to larger distances.
\end{abstract}


\section{Introduction} \label{sec:intro}

Powerful outbursts from supermassive black holes (SMBHs) at the centers of galaxy clusters -- hereafter clusters -- are thought to be the primary drivers that regulate star formation and quench cooling flows. Jets driven by these active galactic nuclei (AGN) {have been observed to} deposit $10^{56}$–$10^{62}$ erg into the intracluster medium (ICM), establishing a feedback loop where cooling gas fuels the AGN, which then heats the ICM (\citealt{McNamaraNulsen2007,2012Fabian,2012Gitti}; but see \citealt{Fabian_hiddend1} for evidence of hidden cooling flows).


Despite a broad consensus on its importance, major uncertainties remain in how the AGN releases energy and how that energy is dissipated in the ICM. A variety of processes may mediate the transport and thermalization of feedback energy \citep[e.g.,][]{2012Gitti}. These include environment-driven mechanisms such as sloshing motions from minor mergers \citep[e.g.,][]{2010ZuHone_slosh_heat}, as well as AGN-driven processes: the inflation of X-ray cavities \citep[also called bubbles, e.g.,][]{2004Birzan}, turbulence induced by jets and/or cavity motions \citep[e.g.,][]{2014Zhuravleva,2022Zhang}, cosmic ray injection and diffusion \citep[e.g.,][]{2007Pfrommer,Sijacki2008,2018Ehlert}, shocks \citep[e.g.,][]{Fabian2006_Perseus_Shock} and sound waves \citep[e.g.,][]{2008Sanders_soundwaves} generated by AGN outflows or from X-ray cavity expansion.

Understanding how efficiently each of these processes couples to the intracluster gas remains a major challenge. While earlier studies often sought a dominant or sole mechanism driving AGN feedback heating, it is increasingly clear that energy deposition results from a combination of several processes, each interacting in complex ways and operating across a range of spatial and temporal scales. For instance, X-ray cavity and AGN-driven shock heating may dominate near and just beyond the radio lobes, sound wave dissipation may prevail at larger radii, thermal conduction from hotter outer layers \citep{2001Narayan,2004Dolag} may become increasingly important farther out \citep{2021Eckert}. 

However, in observational studies, the imprint of AGN feedback is most often detected and studied through X-ray cavities, visible as X-ray dim regions in high-resolution X-ray imaging of nearby clusters (with the \textit{Chandra} X-ray Observatory or \textit{XMM-Newton}). Cavities appear to be present mostly in cool core clusters \citep[$>50\%$ , e.g.,][]{2009Mittal,2012Birzan,HLV2012,2015Hlv,2023Olivares}, which are defined by low central cooling times ($<$7.7Gyr) and entropies ($<$150 kev.cm$^{2}$) and a prominent central X-ray brightness peak due to high gas density \citep[][]{2010Hudson_what_is_a_CC}. These bubbles are thought to be inflated by relativistic plasma during AGN outbursts and appear to be filled by low-frequency synchrotron-emitting plasma launched by the SMBH in the form of jets, providing direct evidence of mechanical energy input. Since X-ray cavities are the most commonly detected feedback-related structures, the AGN energy input is usually gauged from their properties, as enthalpy and power. The latter correlates closely with X-ray cooling losses in the cluster core (the central $\sim$ hundred kpc; e.g., \citealt{2009Cavagnolo}). According to, e.g., \citet[][]{2004Birzan}, X-ray cavities are thought to heat the intracluster gas primarily through the $PdV$ work done during their inflation by the AGN, and through mixing of the hot cavity plasma with the surrounding gas. As cavities rise buoyantly, surrounding gas flows inward to fill their wake, converting gravitational potential energy into kinetic energy, which can then dissipate as heat. However, many uncertainties remain about the cavity contents (hot plasma or cosmic rays \citep{MathewsBrighenti2007,2025Geris_SZ_Cavities}, projection effects limit precise cavity distance and volume estimates \citep{2023Giri}, and the sample selection function of X-ray imagers is biased toward bright and well-studied clusters \citep{2014Panagoulia_vol}. These factors contribute to significant uncertainties in estimating AGN feedback energetics from cavities alone.

While a great deal of attention has been devoted to studying cavities, AGN-driven shock fronts may be another important X-ray feature arising from AGN–ICM interactions \citep[][]{Liu_2019_3C88,2023Ubertosi_rbs}. So far, only about a dozen galaxy groups and clusters have been detected with shocks in their core $<$200 kiloparsecs (kpc) region, which is over an order of magnitude fewer than the number of known X-ray cavities \citep[][]{2016Shin}. The majority of these shock hosts are popular and well-studied systems such as Hydra A \citep[e.g.,][]{2007Wise_HydraA},  Hercules A \citep{Ubertosi2025_HerculesA}, or the galaxy group NGC 5813 \citep[][]{2011Randall_NGC5813}. These observed shocks have often a relatively low Mach number (weak shocks with $\mathcal{M} \sim 1.2$–$2.0$) and can coincide spatially with X-ray cavities and radio lobes \citep[e.g., Cygnus A;][]{Wilson2006_CygnusA} or be elliptical cocoon shocks offsetted from these features \citep[e.g., M~87;][]{Forman2017_M87}. Even fewer massive clusters (M$_{500c}$\footnote{$M_{500\text{c}}$ denotes in a sphere whose mean density is 500 times the critical density of the Universe, at the time the halo is considered.} $> 10^{14} $M$_\odot$) show multiple shock fronts, such as A~2052 \citep{Blanton2009_2052}, and RBS~797 \citep{2023Ubertosi_rbs}. Observationally, shocks appear as sharp arcs in the X-ray surface brightness images, and their detection demands both high spatial resolution (arcsecond) to resolve brightness edges, and high sensitivity, to measure the shallow associated temperature jumps. Their discovery is therefore limited by the long exposures required to detect and characterize them \citep{2008Graham_detceting_sound_waves,Blanton_shock_exposure}.

Consequently, the role of shocks in the AGN feedback cycle remains observationally poorly constrained, though AGN-driven shocks are a promising heating mechanism, capable of thermalizing kinetic energy through gas compression and entropy increase. They are a major science case for proposed next-generation X-ray observatories (e.g., AXIS Science Case Book in prep.). Moreover, since shocks are typically observed as large arcs or elliptical kpc fronts, they are expected to heat the ICM more isotropically than cavities. Estimates of shock power often match \citep[e.g., MS0735+7421;][]{Vantyghem2014_MS} or exceed \citep[e.g., 3C444;][]{Croston2011_3C444} that of X-ray cavities, suggesting that X-ray cavities may account for only about part of the total AGN outburst energy. Therefore, cavity power estimates may represent a lower limit on the AGN’s true energy output.

From a theoretical perspective, many studies have explored the role of AGN-driven shocks in clusters, using both analytical models \citep[such as in][]{2005Voit_theory_agn_heating} and numerical simulations \citep[e.g.,][]{2011Gaspari_Melioli}. Most simulations focus on jet injection in idealized, isolated cluster environments \citep{2014Perucho,2016Yang,2017Weinberger_jet_ICM}, though some incorporate more realistic setups with cosmological realism \citep[][]{2006Heinz,2012Mendygral,2021Bourne_weak_sim}. These studies consistently find that weak shocks and cavities are naturally generated, though their relative contribution to the heating budget varies. For instance, simulations by \citet{2021Bourne_weak_sim} show that weak shocks contribute significantly via gentle, compressive heating of the gas at shock fronts. Lobe-driven sound waves can also decouple and propagate isotropically, offering an additional channel for distributing the AGN energy \citep[e.g.,][]{2004Ruszkowski,Friedman2012_sim}. However, current simulations either focus on a small number of idealized systems or lack a full cosmological environment. As a result, they do not capture the full diversity of AGN-driven shock properties across different systems with various cosmological histories, nor can they robustly assess how common and impactful these shocks are.\\

Given the very limited number of observed clusters with documented AGN-driven shocks (only ten to date; see Table \ref{tab:shocks} and Fig.~\ref{fig:sample_selection_panel}), observational statistics remain sparse. Although these striking examples provide tantalizing insights, a direct and unbiased comparison with simulations is challenging due to the unknown selection function of these observations. This leaves several fundamental questions open: How common are shocks in the population of clusters in general? What are the typical properties of these shock fronts? Since AGN heating {is though to} arise from a complex interplay of jet-driven outflows and X-ray features (cavities, shocks, and pressure waves), to what extent do shocks contribute to the total energy budget? Moreover, how does the dominant heating mechanism vary spatially within clusters?

To address these questions and characterize the population of shocks in galaxy cluster cores within a cosmological context, we use TNG-Cluster\footnote{\url{https://www.tng-project.org/cluster/}} \citep{2024Nelson}, a magnetohydrodynamic cosmological simulation of 352 zoom-in clusters, extending the IllustrisTNG \citep[hereafter TNG,][]{2019NelsonPublicReleaseTNG} project. The simulation suite uses the unchanged TNG model and incorporates a wide range of galaxy and cluster formation and evolution processes, including mergers, gas accretion from the intergalactic medium and ICM outskirts, detailed galaxy formation physics, feedback mechanisms, and magnetic fields. 

This comprehensive model provides a highly realistic environment in which to study AGN-driven signatures on small -- kpc -- scales. {In fact, the} AGN feedback in TNG is implemented as a sub-grid model of intermittent kinetic winds from low-accreting SMBHs -- directed randomly and not fully capturing the microphysics of lobe inflation and ICM interaction. Nonetheless this feedback naturally produces a striking variety of small-scale X-ray structures. These include X-ray cavities that closely match observations \citep[][]{Prunier2025a,Prunier2025b} and eRosita-like bubbles in Milky-Way-like galaxies \citep{2021PillepichErosita}. Only two other recent cosmological simulations also show X-ray cavities arising from different SMBH feedback models: RomulusC \citep{2019Tremmel}, a high-resolution cluster zoom-in from thermal feedback with shut-off cooling, and the Hyenas suite \citep{2024Jennings}, with SIMBA’s bipolar kinetic feedback, producing a diverse cavity population in galaxy groups consistent with observed scaling relations.

In TNG, AGN-driven shocks are found at the edges of eROSITA bubbles, and X-ray cavities in about one-third of cases \citep[][]{Prunier2025a}. As shown by \citet{2017Rainer}, the kinetic feedback model in TNG triggers the quenching of galaxies not only by ejecting gas but also by increasing cooling times through shock heating, hinting at their crucial importance {in the TNG simulations.} However, while the characteristics of AGN-driven shocks remain observationally uncertain due to the limited detected sample, it is equally unclear whether the shocks produced by the TNG model are consistent with observations.

In this work, we present the first systematic study of AGN-driven shocks in a large sample of galaxy clusters drawn from a cosmological simulation, using observational techniques. We generate mock \textit{Chandra} X-ray images of 100 mass-selected halos from the TNG-Cluster suite {and} search for and characterize shock fronts following procedures commonly used in observational data analysis. This approach reveals the diverse, emergent shock morphologies produced by the TNG kinetic feedback model in a fully cosmological context{, without any choices made to reproduce X-ray morphology and without cosmic rays. }It also enables a direct comparison with the small number of observed systems hosting shocks.\\

This paper is organized as follows. In Section~\ref{sec:meth}, we describe the selection of a TNG-Cluster subsample, along with our procedure for identifying and characterizing shocks. Section~\ref{sec:results} presents our main results, the detection, demographics, properties, and energetics of the shock population, and compares them with observational data.  We also use the population of AGN-driven X-ray cavities previously identified in all TNG-Cluster halos \citep{Prunier2025a} to characterize and quantify the X-ray cavity population within in the same TNG-Cluster subsample. In Section~\ref{sec:discussion}, we interpret our results and their implications before concluding in Section~\ref{sec:conclusion}.

\section{Method}\label{sec:meth}

First, we describe the TNG-Cluster cosmological simulation and its diverse AGN feedback-driven X-ray features in subsection~\ref{subsec:meth_tng-cluster}. Next, we present TNG-Cluster subsample, matched in mass to the ten observed clusters gathered in Table \ref{tab:shocks},{ but ten times larger (100 halos),} in subsection~\ref{subsec:meth_sample_sel}. We then describe the mock \textit{Chandra} observation procedure in \ref{subsec:mock} and the shock detection and characterization pipeline in \ref{subsec:meth_shock_prop}. Finally, we outline the procedure to compute cavity power \ref{subsec:cav} and cooling luminosities \ref{subsec:met_cool_lum}.

\subsection{TNG-Cluster, SMBH feedback model, and X-ray morphological features}\label{subsec:meth_tng-cluster}

TNG-Cluster \citep{2024Nelson} is a set of 352 zoom-in simulations of massive galaxy clusters simulated with the TNG magnetohydrodynamic galaxy formation model \citep[][and references therein]{2017Rainer,2018Pillepich,2019NelsonPublicReleaseTNG}. Clusters were selected from a parent, dark matter-only, simulation box of 1 Gpc$^3$, focusing on cluster mass range $M_{500\text{c}}$ = 10$^{14-15.4}$ M$_\odot$ at $z=0$, and re-simulated with the baryonic mass resolution of TNG300 ($m_\text{gas} \sim 10^{7}$\msun), which is 30–100 times higher than that of comparable large-volume simulations \citep{2018Cui_three_hundred,2017McCarthy_BAHAMAS}. Moreover, the adaptive Voronoi mesh in AREPO provides an average hydrodynamic spatial resolution in cluster cores (within $<0.5$ R$_\text{500c}$) ranging from 1 to 8 kpc, with individual cells sometimes reaching sizes below 1 kpc in high density regions, allowing the simulations to capture small-scale structures with high fidelity.

The TNG model{, adopted unchanged in TNG-Cluster, }includes gas cooling and heating, star formation, metal enrichment, stellar feedback, SMBH seeding and growth, and AGN feedback (without cosmic rays). SMBHs grow via Bondi accretion and mergers, releasing feedback energy either in thermal or kinetic forms, alongside continuous radiative-like feedback \citep{2013Vogelsberger}. At high accretion rates, SMBHs inject thermal, isotropic energy (quasar mode), while at low rates, feedback shifts to kinetic, directional outbursts (radio mode), delivering momentum without immediate heating  \citep{2017Rainer}. The delivered energy is parametrized as $\Delta \dot{E}_{\text{kin}} = \epsilon \dot{M}_{\text{SMBH}} c^2$, with a maximum coupling efficiency of $\epsilon = 0.2$ \citep{2017Rainer,2018Pillepich}. We measure the typical power of feedback event {in clusters} of $10^{41} - 10^{45} \, \text{erg s}^{-1}$ at low-redshifts \citep{Prunier2025a}, reaching even higher values in lower mass galaxies \citep[][]{2021PillepichErosita}. This subgrid approach, rather than modeling narrow, collimated jets, approximates the large-scale effects of AGN feedback within cosmological resolution limits. It is physically motivated by high-velocity accretion-disk winds or compact jets from low-luminosity SMBHs \citep{2014Yuan}, effectively quenches star formation in massive galaxies via strong outflows \citep{2019Nelsonb} {and offsets cooling of the halo gas \citep{zinger2020ejective-c4c}. }

A wealth of AGN-driven X-ray morphological structures emerges naturally in the simulations due to this kinetic SMBH feedback. These include eROSITA-like bubbles in Milky Way-mass galaxies \citep{2021PillepichErosita} and X-ray cavities in the brightest cluster galaxies (BCGs), shocks, and pressure waves, \citep[e.g.,][]{2024Truong,Prunier2025a} demonstrating the model’s ability to reproduce feedback-induced features despite not explicitly modeling jet physics. In \citet{Prunier2025a} and \citet{Prunier2025b}, we showed that the population of X-ray cavities {in TNG-Cluster} is realistic and diverse, characterized by underdense X-ray dim regions filled with hot gas ($\sim10^8$ K), typically in pressure equilibrium with the surrounding ICM. These cavities span a wide range of morphologies and evolutionary stages, including single, paired, and multiple cavities within the same system. About 25 per cent are surrounded by X-ray bright, compressed rims often associated with weak shocks (Mach number < 2, from the shock finder \citep[][]{2016Schaal}. When compared carefully to observed X-ray cavities, TNG-Cluster cavities show comparable sizes and morphologies (though cavities smaller than 10 kpc are underrepresented), and occupy the same loci in scaling relations such as cavity distance versus area and cavity power versus cooling luminosity \citep[][]{Prunier2025b}. These results support the idea that X-ray cavities in the simulation play a significant role in heating cluster cores.

\subsection{Observed clusters with AGN-driven shocks and mass-matched selection of TNG-Cluster analogs}\label{subsec:meth_sample_sel} 

To date, fewer than ten galaxy clusters \footnote{We decide on a mass threshold between groups and galaxy clusters of M$_{500c}$ $> 10^{14} $M$_\odot$ \citep[][]{2012Kravtsov}.} with confirmed AGN-driven shock signatures have been reported in the literature (see Table~\ref{tab:shocks}); all are located within the central $\sim$ 200 kpc of cool core clusters and exhibit clear evidence of AGN-driven origins, either through spatial coincidence with X-ray cavities or cocoon-like structures surrounding radio lobes. In contrast, merger shocks are more commonly observed in dynamically disturbed, non-cool core systems and typically differ in key properties such as higher Mach numbers (> 2), larger distances from cluster centers, and distinct morphologies (\citealt{2016Eckert,2019Golovich} and see Discussion \ref{sec:discussion_represent}).

The discovery of AGN-driven shocks in galaxy clusters is limited due to the long X-ray exposure times and high spatial resolution typically required to detect such features, but also because of the characteristic weakness of these shocks. The ten clusters with detected AGN-driven shocks invariably host X-ray cavities, often multiple -- up to six in Hydra A -- and exhibit one or more shocks with Mach numbers ranging from 1.1 to 1.9. These shocks display a variety of morphologies, including elliptical arcs or cocoon (i.e., closed ellipse) fronts, as seen in A2052 \citep[e.g.,][]{2011Blanton_A2052} and M87 \citep[e.g.,][]{Forman2017_M87}, as well as shocks located at the boundaries of X-ray cavities, as in Cygnus A \citep[e.g.,][]{Wilson2006_CygnusA}. Some clusters have both, such as RBS 797 \citep[e.g.,][]{2023Ubertosi_rbs}, which features an inner shock encompassing the central cavities and four arc-shaped shock fronts at larger radii.

The ten reported clusters listed in Table~\ref{tab:shocks} are well-studied systems, each with multiple publications and deep \textit{Chandra} and/or \textit{XMM-Newton} coverage across several observing cycles \citep[e.g.,][]{2007Wise_HydraA,2009Simionescu_HydraA}. Notably, often these clusters were initially targeted due to the presence of X-ray cavities, with subsequent deeper observations later revealing shocks. This introduces a possible strong selection bias: AGN-driven shocks are more likely to be found in systems already known to host strong AGN feedback features. As a result, this limited sample of ten observed clusters with detected shocks is likely not representative of the broader population.

Our present study of shocks in the TNG-Cluster simulation is not intended to reproduce the potential underlying selection bias inherent in the observational sample. Instead, we adopt an agnostic approach to shock detection while restricting our analysis to a similar mass regime. {In fact, }X-ray cavity properties such as powers are known to correlate with systems' mass, with on average smaller powers in groups \citep{2012Gitti}. 

Therefore, we select a subsample of 100 simulated clusters solely based on mass to match the mass range of the observed sample ($M_{500\text{c}}$ between $1.2$ and $8.5 \times 10^{14}\ \mathrm{M}_\odot$). We randomly draw these 100 clusters from the 294 TNG-Cluster halos out of 352 with $M_{500\text{c}}$ in that range. Compared to the full TNG-Cluster population, our sample does not include the highest-mass clusters above $10^{15} \, M_{\odot}$.

\begin{figure}
\centering
    \includegraphics[width=0.55\textwidth]{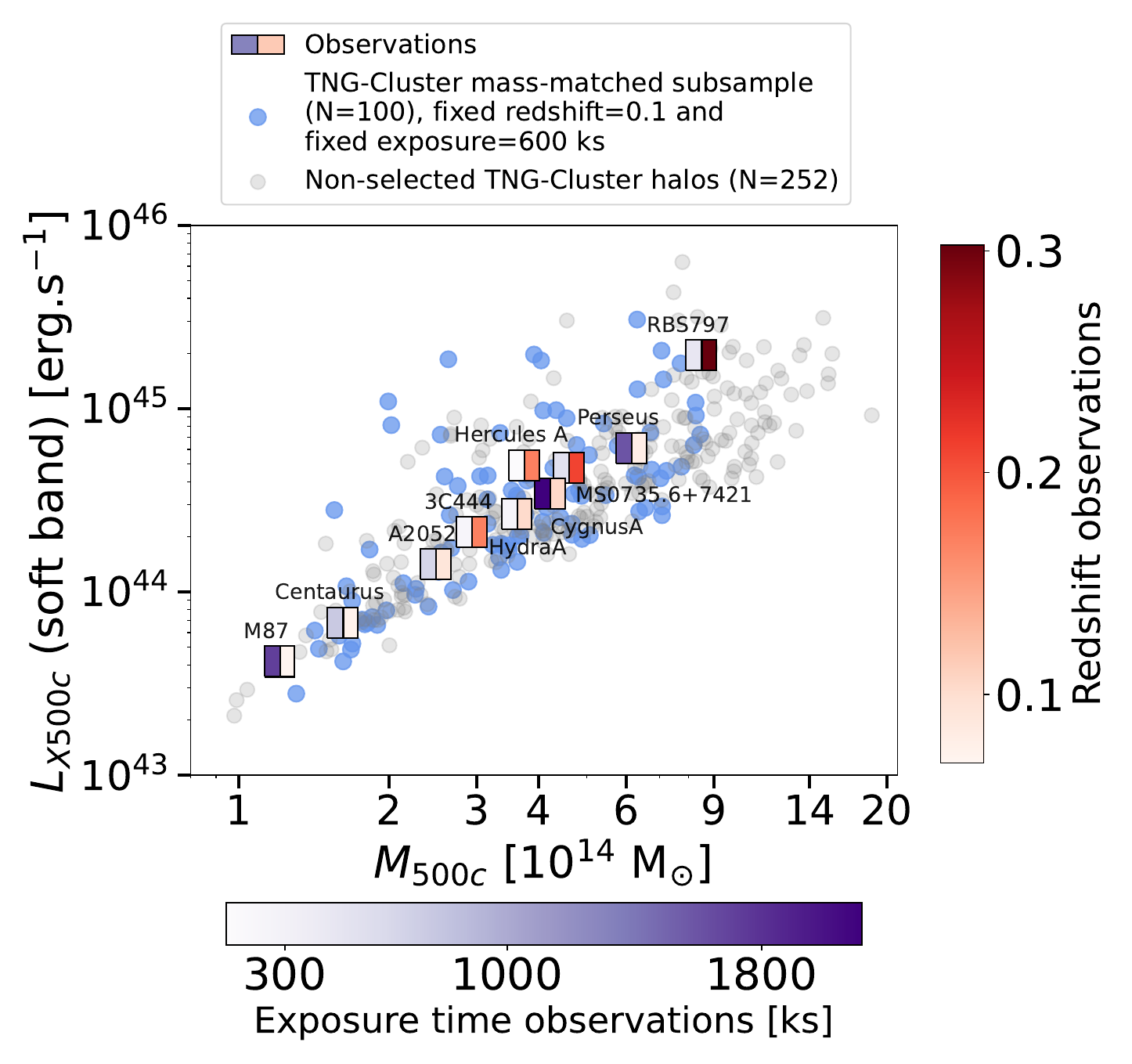}
    \vspace{-0.5cm}
    \caption{TNG-Cluster mass-selected sample of 100 {simulated clusters from TNG-Cluster} (blue circles) randomly chosen to broadly match the M$_\text{500c}$ of the 10 known observed clusters hosting shocks (Table \ref{fig:sample_selection_panel}{, rectangle markers}). Each observational point is color-coded by its total \textit{Chandra} exposure time (left) and redshift (right). Although the TNG-Cluster subsample was selected based solely on mass, it exhibits soft X-ray luminosities broadly consistent with observations -- except for a small subset with slightly higher L$_\text{x500c}$  -- and is representative of the full TNG{-Cluster} population within this mass range.}
    \label{fig:sample_selection_panel}
\end{figure}

Figure \ref{fig:sample_selection_panel} shows the chosen TNG-Cluster halos (blue) against the overall {TNG-Cluster} population (gray) in the space defined by X-ray luminosity $L_{\mathrm{X,500\text{c}}}$ and mass M$_{500\text{c}}$, along with the observed ten clusters color-coded by total exposure time and redshift. For the simulated clusters, the intrinsic X-ray luminosity corresponds to the soft band (0.5–2.0 keV) luminosity derived using an APEC model in post-processing \citep{2024Nelson}, measured within R$_{500\text{c}}$. For the observed cluster,  M$_{500\text{c}}$ and $L_{\mathrm{X,500c}}$ in the 0.1–2.4 keV band within R$_{500\text{c}}$, are quoted from the MCXC catalog \citep{2011Piffaretti_catalog_mcxc}.

The selected TNG-Cluster halos exhibit similar properties and follow a similar  $L_{\mathrm{X,500\text{c}}}$ - M$_{500\text{c}}$ trend as the observed sample of ten clusters, as illustrated in Figure~\ref{fig:sample_selection_panel}. We note that a small subset ($\sim$8 out of 100 clusters), lies about 0.3 dex above the best-fit observational relation in logarithmic space, indicating a mild positive offset in $L_{\mathrm{X,500c}}$ at fixed mass.

\begin{table*}
    \centering
    \caption{Properties of clusters hosting both X-ray detected shocks and cavities with available observation(s). (1) Source name, (2) total mass within $R_{500\text{c}}$, (3) redshift, (4) total exposure time with \textit{Chandra} ACIS-I, (5) number of cavities, (6) number of shocks, (7) Mach numbers of detected shocks, (8–10) cavity, shock, and total AGN power in units of $10^{43}$ erg s$^{-1}$, (11) reference. Values enclosed in square brackets [ ] indicate the range or uncertainty of the quantity when a single value was not available.
    [1] \citet[][]{2023Ubertosi_rbs},[2] \citet{Graham2008_perseusWS}, [3] \citet{Vantyghem2014_MS}, [4] \citet[][]{2021Biava}, [5] \citet{Nulsen_2005_HydraA}, [6] \citet{2007Wise_HydraA}, [7] \citet{2012Birzan}, [8] \citet{snios2018cocoon-a43} [9] \citet{Blanton2009_2052}, [10] \citet{Croston2011_3C444}, [11] \citet{cav_2017Vagshette}, [12] \citet{2016Sanders_centaurus}, [13] \citet{2001churazov}, [14] \citet{2005Forman}, [15] \citet{2007Forman}, [16] \citet{Forman2017_M87}, 
     [17] \citet{2005Nusen_HerculesA_sh_cav}, [18] \citet{2012Gitti}, [19] \citet{Ubertosi2025_HerculesA}, [20] \citet[][]{2004Birzan}, [21] \citet{2006Rafferty}}
    \label{tab:shocks}
    \begin{tabular}{lccccccccccc}
    \hline
    \\
    Source  & $M_{500\text{c}}$ & $z$ & $t_{\mathrm{exp}}$  & $N_{\mathrm{cav}}$ & $N_{\mathrm{shock}}$ & Mach & $P_{\mathrm{cav}}$ & $P_{\mathrm{shock}}$ & $P_{\mathrm{tot}}$ & Ref\\ 
           & [$10^{14} M_\odot$] & & [ks] & & & & [$10^{43}$ erg s$^{-1}$] & [$10^{43}$ erg s$^{-1}$] & [$10^{43}$ erg s$^{-1}$] & \\
          (1) & (2) & (3) & (4) & (5) & (6) & (7) & (8) & (9) & (10) & (11) \\
    \\
    \hline
    \\
    RBS 797 & 8.50 & 0.354 & 427 & 4 & 3 & 1.20, 1.19, 1.22 & 360 & 2900 & 3360 & 1 \\
    Perseus & 6.15 & 0.0179 & 1594 & 4 & 1 & 1.21 & 32 & 120 & 152 & 2, 20\\
    MS 0735+74 & 4.61 & 0.215 & 527 & 4 & 1 & 1.26 & 1700 & 1100 & 2800 & 3, 4, 20 \\
    Cygnus A & 4.22 & 0.0561 & 2117 & 2 & 2 & 1.18, 1.18 & 210 & 1000 & 1210 & 7, 8, 20 \\
    Hydra A & 3.62 & 0.0539 & 239 & 6 & 1 & 1.2 & 210 & 200 & 410 & 5, 6\\
    3C 444 & 2.94 & 0.153 & 218 & 2 & 1 & 1.3 & [61-127] & 2900 & [2961- 3027] & 10,  11 \\
    A~2052 & 2.49 & 0.0353 & 654 & 4 & 2 & 1.18, 1.20 & 3.2 & 1.0 & 4.2 & 9, 20 \\
    Centaurus& 1.62 & 0.0104 & 760 & 2 & 1 & [1.1-1.4] & 1.2 & 0.74 & 1.94 & 12, 20, 21 \\
    M87 & 1.21 & 0.0036 & 1720 & 2 & 2 & 1.2, 1.22 & 1.0 & 2.4 & 3.4 & 13, 14, 15, 16, 20 \\
    Hercules A & 3.75 & 0.155 & 115 & 2 & 2 & 1.65, 1.9 & 31 & [1400-2000] & [1431- 2031] & 17, 18, 19 \\
    \\
    \hline
    \end{tabular}
\end{table*}

\subsection{Mock \textit{Chandra} observations of simulated {clusters}} \label{subsec:mock} 
Once selected, we produce mock Chandra X-ray observations of each of the 100 TNG-Cluster selected halos using the pyXSIM \citep{pyXSIM_ZuHone} and SOXS \citep{SOXS_ZuHone} software suites. For each gas cell within a cubic region of $\pm2\times$R$_{500}$ centered on the SMBH of the BCG (i.e., the most massive galaxy within the friends-of-friends halo of each cluster), we generate a mock X-ray spectrum based on its density, temperature, and metallicity, assuming simulation abundance ratios, a single-temperature APEC model, and galactic absorption with a hydrogen column density of $N_{\mathrm{H}} = 3\times10^{20}$ cm$^{-2}$. The resulting spectra are summed to create a large sample of intrinsic photons, which are then projected onto the detector plane and convolved with the Cycle 19 ACIS-I instrumental response and on-axis point spread function to produce mock \textit{Chandra} event files. These mock observations include instrumental backgrounds and the galactic foreground and are limited to the 0.5–7.0 keV energy band to match the ACIS broad energy range. Exactly one mock is generated for each cluster {, with the projection arbitrarily chosen along the simulation box z-axis.} The redshift range of our observed cluster sample spans from 0.0036 to 0.3540, and has a median exposure time of 590 ks. Therefore we fix the exposure time to 600 ks for the mocks, and to ensure enough spatial resolution while keeping a large field of view for the shock detection, we fix the cluster distance to 460 Mpc (corresponding to $z = 0.1$), yielding a \textit{Chandra} resolution of 0.938 kpc per pixel and an ACIS-I field of view of $\sim$1.5 Mpc.

\subsection{Detecting and measuring shock properties} \label{subsec:meth_shock_prop}
In the following, we outline the procedure used to identify and characterize shock fronts. Detecting shocks in the X-rays requires to (i) detect visually an edge in surface brightness in X-ray images, (ii) determine the precise location and strength of each front through surface brightness profiles, and (iii) spectroscopically measuring thermodynamic properties (e.g., temperature and pressure) on either side of the front to confirm the shock nature and estimate its power. Our approach is based on methods developed in previous studies of individual observed systems with shocks \citep[e.g.,][]{Gitti2011_HydraA,2023Ubertosi_rbs}. The key difference is that we apply this procedure to an unprecedentedly large sample of 100 mock observations, often featuring multiple shock candidates per image. To keep the analysis tractable, we streamlined the workflow and omitted optional steps which are sometimes applied in single-object studies, such as full 2D temperature mapping, or comparing different deprojection algorithms \citep{Randall2011_NGC5813_Group,2023Ubertosi_rbs}, prioritizing efficiency and consistency across the sample.

\paragraph*{Detection.}
We identify candidate shocks by visually inspecting the mock \textit{Chandra} X-ray images. 
To aid in their identification, we apply spatial frequency filtering techniques to the images, namely unsharp masking (UM) and the Gaussian Gradient Magnitude (GGM) filter \citep{2016Sanders,2021ascl.soft03006S}. UM enhances contrast by subtracting a smoothed version of the image from the original, highlighting small-scale features like shock fronts. The GGM filter computes the gradient of the smoothed image, effectively tracing surface brightness edges. Both filters are applied with varying Gaussian widths to capture features at different spatial scales (see Figure \ref{fig:panel_shocks}). This by-eye identification process is a first step to localize candidate shock regions in each image. For each cluster, we focus on up to four of the most prominent shock candidates to limit the load of the follow-up analysis (most of the time, fewer than four candidates were effectively detected). For each candidate shock, we create an associated \texttt{Ciao} Panda region, corresponding to wedge-shaped annular sectors defined by radial and azimuthal boundaries, at its approximate location.\\

\paragraph*{Surface brightness profile.} The next step involves extracting surface brightness profiles from the previously identified candidate shock wedge regions and using various binning (between $0.5^{\prime\prime}~\text{to}~1.8^{\prime\prime}$). Each profile is fitted with both a single and a broken power-law model using the \texttt{Proffit} package \citep{2020Eckert}. The broken power-law model includes five parameters: normalization $A_s$, density jump $J$, jump radius $r_J$, and the inner and outer slopes $\alpha_1$ and $\alpha_2$. A shock is considered detected at radius $r_J$ if the broken power-law fit significantly improves over the single power-law (F-test confidence > 99\%). The Mach number $\mathcal{M}$ is then derived from the best-fit density jump $J$ using the Rankine–Hugoniot conditions \citep[e.g.,][]{2007Markevitch}.

\begin{equation}
    \mathcal{M} = \left( \frac{3J}{4 - J} \right)^{1/2}
\end{equation}

Figure \ref{fig:panel_shocks_fit_temp_jump} shows several examples of surface brightness profiles with their best fits with broken power-law models. Although not shown here, 4 out of 60 shock candidates {associated with} deep X-ray cavities {exhibit} depressions (downward turns) in the surface brightness profile at their locations. In these cases, fitting a broken power-law was challenging, and the fit might underestimate the brightness jump, despite the clear visual front seen in the images. In one instance, we were unable to obtain a reliable fit.

\paragraph*{Temperature and density before and after the shock.}

A shock is thermodynamically characterized by a jump in both temperature and density between the pre- and post- shock regions, where the gas has been heated and compressed by the shock passage. Detecting these jumps immediately before and after the front is essential for two reasons: (i) to confirm the shock nature, since a cold front can produce similar surface brightness discontinuities but with an inverse temperature gradient, and (ii) to compute the pressure jump required to estimate the shock power (see paragraph below).

For each shock candidate, we {extract} spectra from three wedge regions. The first region spans from 0.6-0.8 $\times$ r$_\mathrm{J}$ to r$_\mathrm{J}$ (post-shock), the second spans from r$_\mathrm{J}$ to 1.2-1.6 $\times$ r$_\mathrm{J}$ (pre-shock). These bin widths {are} chosen to stay close to the shock while keeping enough counts for fitting. Larger regions would boost signal-to-noise but risk smoothing out the temperature and density gradients. The third outer region extends from 1.6 $\times$ r$_\mathrm{J}$ out to 3-4$\times$ r$_\mathrm{J}$, and provides a baseline for estimating foreground and background emission, enabling accurate deprojection of the inner regions.

Spectra from these three regions {are} jointly fitted using the CIAO Sherpa Deproject tool, which recovers three-dimensional source properties from two-dimensional X-ray spectra. The fits are performed over the 0.5–7.0 keV range using XSPEC with a single-temperature plasma model and foreground absorption, modeled as \texttt{tbabs*apec}, and evaluated with the Cash statistic. The free parameters {are} temperature, abundance, and normalization, while absorption and redshift {are} fixed. Shock candidates {are} confirmed when a temperature jump of at least 2$\sigma$ significance was detected. In our sample, 8 out of 60 shock candidates were reclassified as cold fronts during this procedure.

\paragraph*{Shock energetics.}
Measuring the age and energetics of the shocks provides essential information for probing the impact of these features on the cluster's thermodynamic conditions. To compute the energy deposited by each shock, we follow the method outlined by e.g., \cite{2011Randall_NGC5813}, \begin{equation}
    P_{\mathrm{shock}} = \frac{E_{\mathrm{shock}}}{t_{\mathrm{age}}} = \frac{1}{t_{\mathrm{age}}}  \frac{3}{2} V \times (p_{\mathrm{in}} - p_{\mathrm{out}}) ,
\end{equation}

where $V$ is the volume of shocked gas i.e. the portion of ellipsoid encompassed by a shock front. The pressure difference corresponds to the pressure jump across the shock front, with $p_{\mathrm{in}}$ and $p_{\mathrm{out}}$ being the downstream and upstream pressures obtained from spectral analysis of the shock regions with \begin{math}\label{eq:pth}
p = (\mu_e/\mu) n_e k_B T
\end{math}, where we assume $\mu_e/\mu$ = 1.92 with $\mu_e$ the mean molecular weight per electron. We estimate the shocked volume assuming a prolate ellipsoidal geometry within the angle defined during the shock identification phase. The shock age, $t_{\rm age}$, can be determined by assuming that the shock has traveled from the center {of the cluster} to its current position, $r_{\rm shock}$ (the distance to the mid-aperture of the front), at its observed Mach number $\mathcal{M}$, i.e.,
\begin{equation}
    t_{\rm age} = \frac{r_{\rm shock}}{\mathcal{M} \, c_s},
\end{equation}
where $c_s = \sqrt{\gamma kT / (\mu m_p)} \approx 5.2 \times 10^2 \sqrt{kT~[{\rm keV}]}$ km~s$^{-1}$ is the upstream sound speed, measured from the pre-shock temperature. We note that this method may slightly overestimate the true shock age (by a relatively modest factor of $\sim$10--20\%; see \citealt{2011Randall_NGC5813}), as the shock likely had a higher Mach number at launch. However, this may be partially compensated by the underestimation of the true shock distance due to projections effects.

\subsection{X-ray cavities energetics}\label{subsec:cav} 
For the subsample of 100 TNG-Cluster mass-selected systems, we also investigate the energetics of X-ray cavities. The full population of X-ray cavities across all 352 TNG-Cluster halos {was} identified in \citet{Prunier2025a} using mock \textit{Chandra} ACIS-I observations with an exposure time of 200 ks{, with the projection chosen along the simulation box z-axis, as in the present work} (these mock \textit{Chandra} event files have been made publicly available). Projected cavity volumes and distances were estimated using observational techniques in \citet{Prunier2025a}, {and}, using these measurements, we compute the associated cavity power, $P_{\mathrm{cav}}$, following methods used in observational studies such as \citet{2014Panagoulia_vol}. Assuming cavities are in approximate pressure equilibrium with the surrounding ICM, the cavity power is computed as \begin{math} P_{\mathrm{cav}} = \frac{E_{\mathrm{cav}}}{t_{\mathrm{cav}}} = \frac{\gamma}{\gamma - 1} \frac{p_{\mathrm{th}} V}{t_{\mathrm{cav}}} \end{math} where $p_{\mathrm{th}}$ is the local thermal pressure of the ICM at the cavity height, $V$ is the cavity volume assuming a prolate ellipsoid geometry, and $t_{\mathrm{cav}}$ is the cavity age. In the TNG simulations, we assume an adiabatic index of $\gamma = 5/3$, appropriate for a non-relativistic monoatomic gas{:} this differs slightly from the typical observational assumption of $\gamma = 4/3$, which reflects the relativistic content of the plasma thought to fill real cavities. The cavity age, $t_{\mathrm{cav}}$, is estimated as the sound crossing time. The procedure is similar to \cite{Prunier2025b} and further details can be found in the method section.

\subsection{Cooling radius and luminosity}\label{subsec:met_cool_lum}
The total X-ray cavity and shock power can be compared with the cluster's gas luminosity inside the cooling radius, $r_{\mathrm{cool}}$, {to get an idea of how much} AGN mechanical feedback {is needed} to prevent runaway cooling (according to the classical `cooling flow problem'). Beyond this, the observed scaling between jet power and X-ray luminosity supports a scenario of self-regulated AGN feedback, in which cooling and heating are energetically balanced over time \citep[e.g.][]{2021Eckert,2022DonahueVoit}.

$L_{\mathrm{cool}}$ is typically defined as the total X-ray luminosity within the region where the gas cooling time is less than $7.7 \times 10^9$ years \citep{2008Birzan,2013McDonalds}, although alternative thresholds, such as 3 Gyr, have been adopted in the literature \citep[e.g.,][]{2014Panagoulia_vol,2018McDonald}. To determine r$_\mathrm{cool}$, we first compute the deprojected radial profile of the cooling time $t_{\mathrm{cool}}$:
\begin{math}
    t_{\mathrm{cool}} = \frac{3 p_{\mathrm{th}}}{n_e n_{\mathrm{H}} \Lambda(Z, T)} = \frac{3 p_{\mathrm{th}} V}{2 L_X},
\end{math}
where $\Lambda(Z, T)$ is the cooling function for gas at metallicity $Z$ and temperature $T$, $p_{\mathrm{th}}$ and $L_X$ are respectively the pressure and the X-ray luminosity measured in each radial annulus of the profile \citep[e.g.,][]{2017Hogan}. We then identify the radius at which $t_{\mathrm{cool}} < 7.7$ Gyr and compute L$_{\mathrm{cool}}$, the bolometric luminosity between 0.001 and 100 keV inside r$_\text{cool}$, by integrating the X-ray luminosity within this aperture on our mock \textit{Chandra} event files. 

\begin{figure*}
    \vspace{-0.3cm}
    \centering
    \includegraphics[trim=0cm 0cm 0cm 1.1cm, clip, width=0.975\textwidth]
    {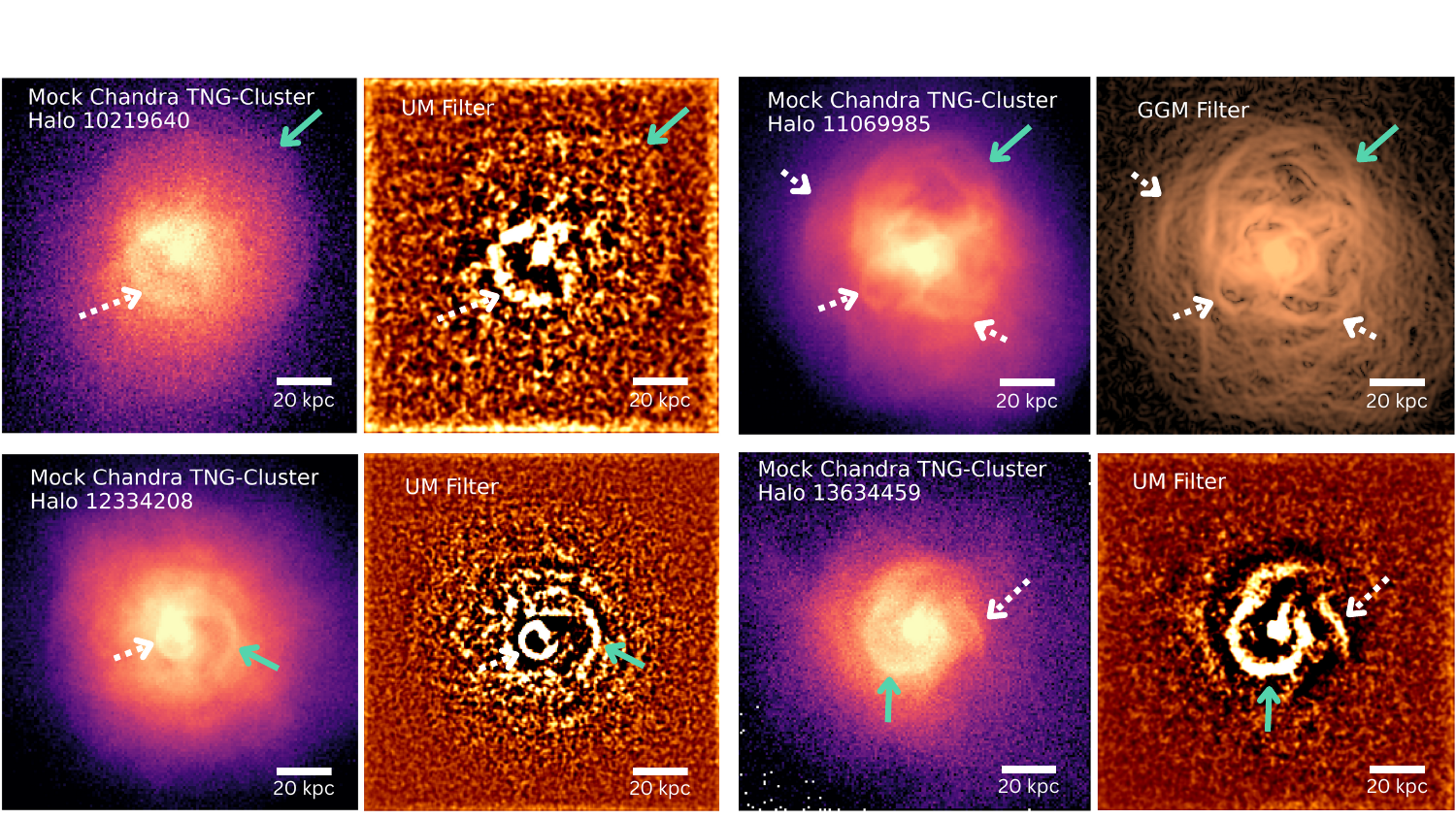}
    \includegraphics[width=0.98\textwidth]{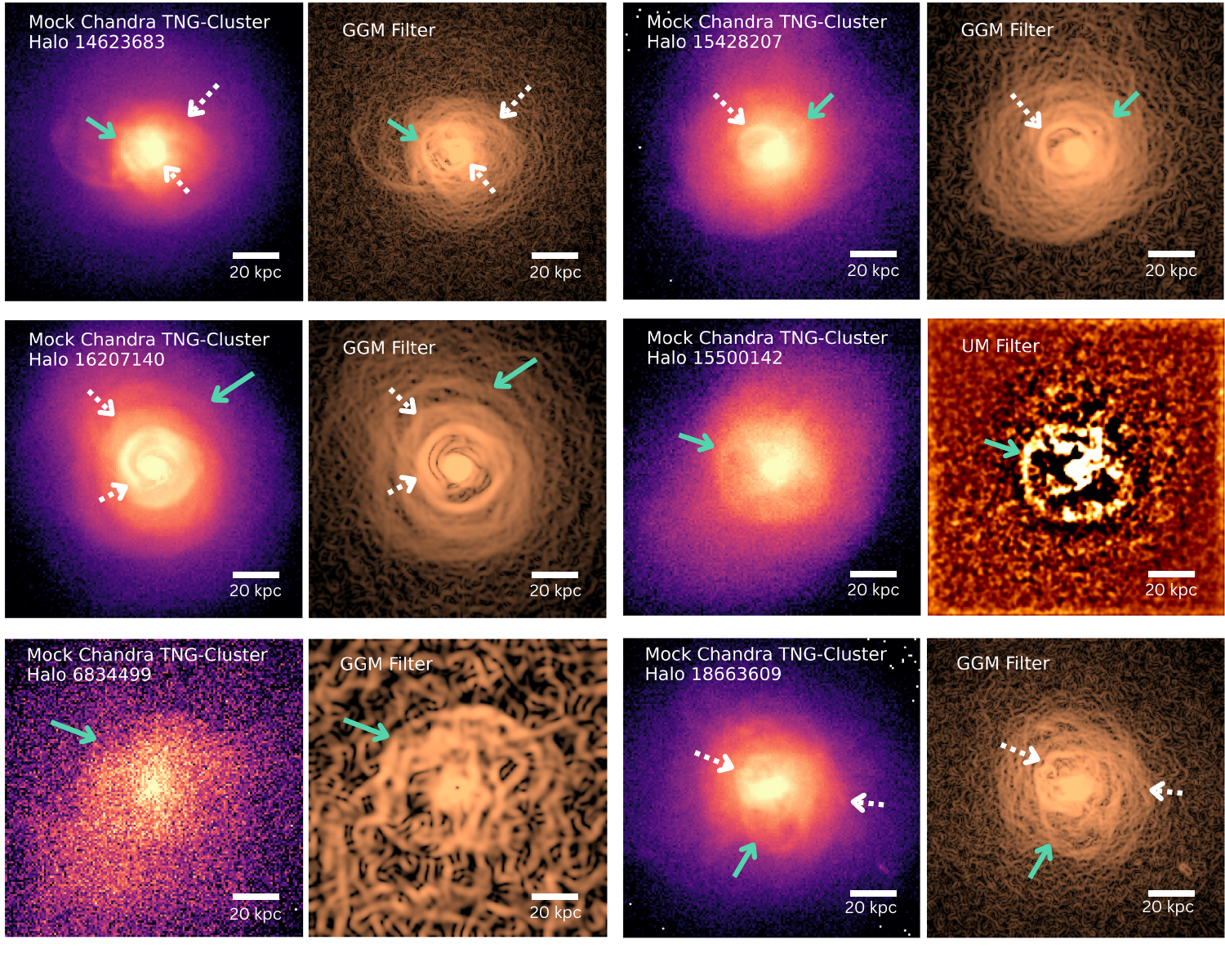}
    \vspace{-0.7cm}
    \caption{Gallery of {simulated galaxy clusters from TNG-Cluster} exhibiting X-ray shocks. Mock \textit{Chandra} surface brightness images of the central $200\times200$ kpc region of six selected clusters in TNG-Cluster, alongside images processed with unsharp masking or Gaussian gradient magnitude filters to highlight the shock fronts. Images are centered on the AGN position. White dashed arrows mark identified shock fronts, while the green arrow points the specific shock for which we show, in Fig.~\ref{fig:panel_shocks_fit_temp_jump}, the surface brightness profile and associated temperature jump. The gallery illustrates the diversity of AGN-driven shocks in TNG-Cluster cores, including features located at cavity edges or at larger radii, and exhibiting a range of morphologies, from arc-like fronts with varying opening angles to complete ellipses.}
    \label{fig:panel_shocks}
\end{figure*}

\begin{figure*}
    \centering
    \begin{minipage}[t]{0.49\textwidth}
        \centering
        \includegraphics[width=\linewidth]{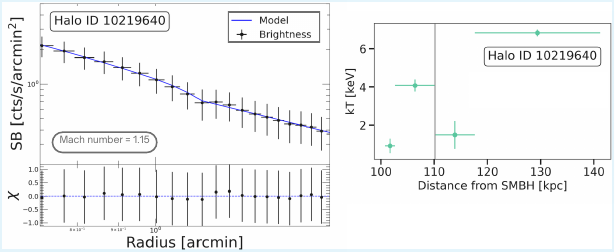}
    \end{minipage}
    \hfill
    \begin{minipage}[t]{0.49\textwidth}
        \centering
        \includegraphics[width=\linewidth]{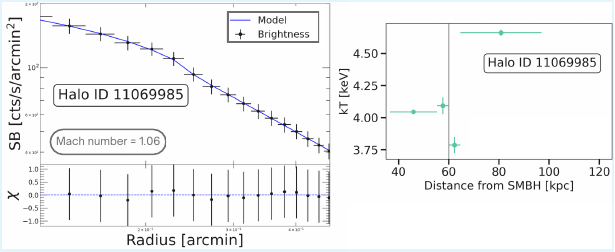}
    \end{minipage}
    
    \vspace{0.5em}
    
    \begin{minipage}[t]{0.49\textwidth}
        \centering
        \includegraphics[width=\linewidth]{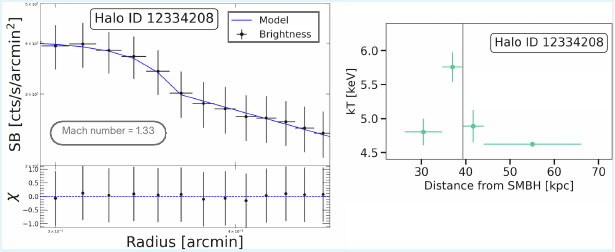}
    \end{minipage}
    \hfill
    \begin{minipage}[t]{0.49\textwidth}
        \centering
        \includegraphics[width=\linewidth]{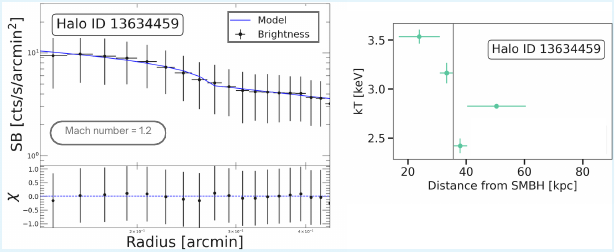}
    \end{minipage}

    \vspace{0.5em}

    \begin{minipage}[t]{0.49\textwidth}
        \centering
        \includegraphics[width=\linewidth]{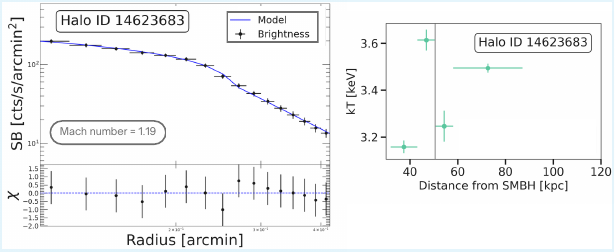}
    \end{minipage}
    \hfill
    \begin{minipage}[t]{0.49\textwidth}
        \centering
        \includegraphics[width=\linewidth]{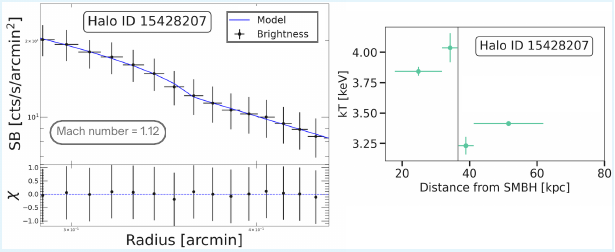}
    \end{minipage}
    
    \vspace{0.5em}
    
    \begin{minipage}[t]{0.49\textwidth}
        \centering
        \includegraphics[width=\linewidth]{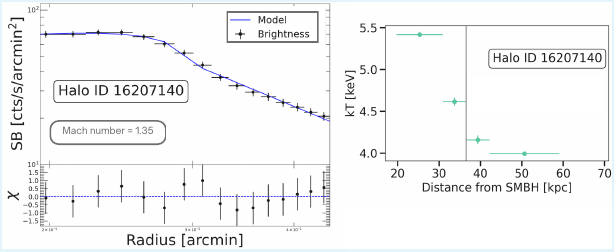}
    \end{minipage}
    \hfill
    \begin{minipage}[t]{0.49\textwidth}
        \centering
        \includegraphics[width=\linewidth]{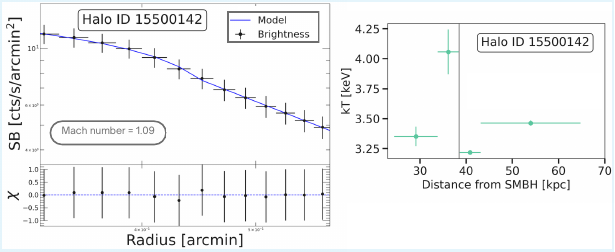}
    \end{minipage}
    
    \vspace{0.5em}
    
    \begin{minipage}[t]{0.49\textwidth}
        \centering
        \includegraphics[width=\linewidth]{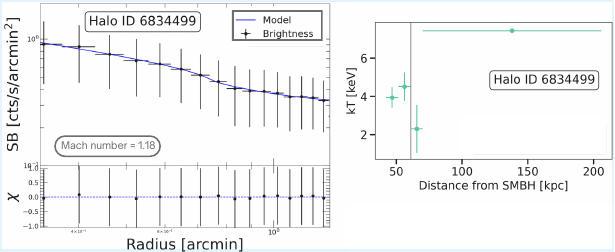}
    \end{minipage}
    \hfill
    \begin{minipage}[t]{0.49\textwidth}
        \centering
        \includegraphics[width=\linewidth]{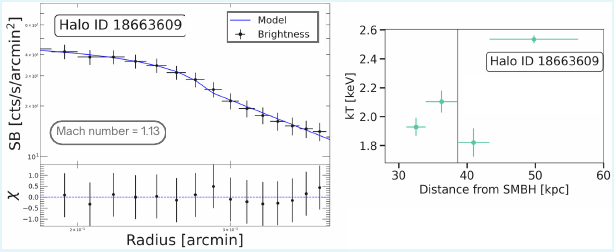}
    \end{minipage}
    \caption{{Profiles across selected shock fronts from the TNG-Cluster simulations} (blue arrows in the {figure} above). {\textit{Left}: Surface brightness profiles at the shock locations, with the quoted Mach numbers inferred from the best-fit broken power law. \textit{Right}: Temperature profiles across the same shock fronts, showing the pre-shock (colder) and post-shock (hotter) temperature jumps.}}
    \label{fig:panel_shocks_fit_temp_jump}
\end{figure*}

\section{Results}\label{sec:results}

\subsection{How common are shocks in the cores of TNG-Cluster halos?} \label{subsec:morpho}

\begin{figure*}
  \centering
   \hspace{-1.8cm}
  \begin{subfigure}[b]{0.52\textwidth}
    \includegraphics[width=1.08\textwidth]{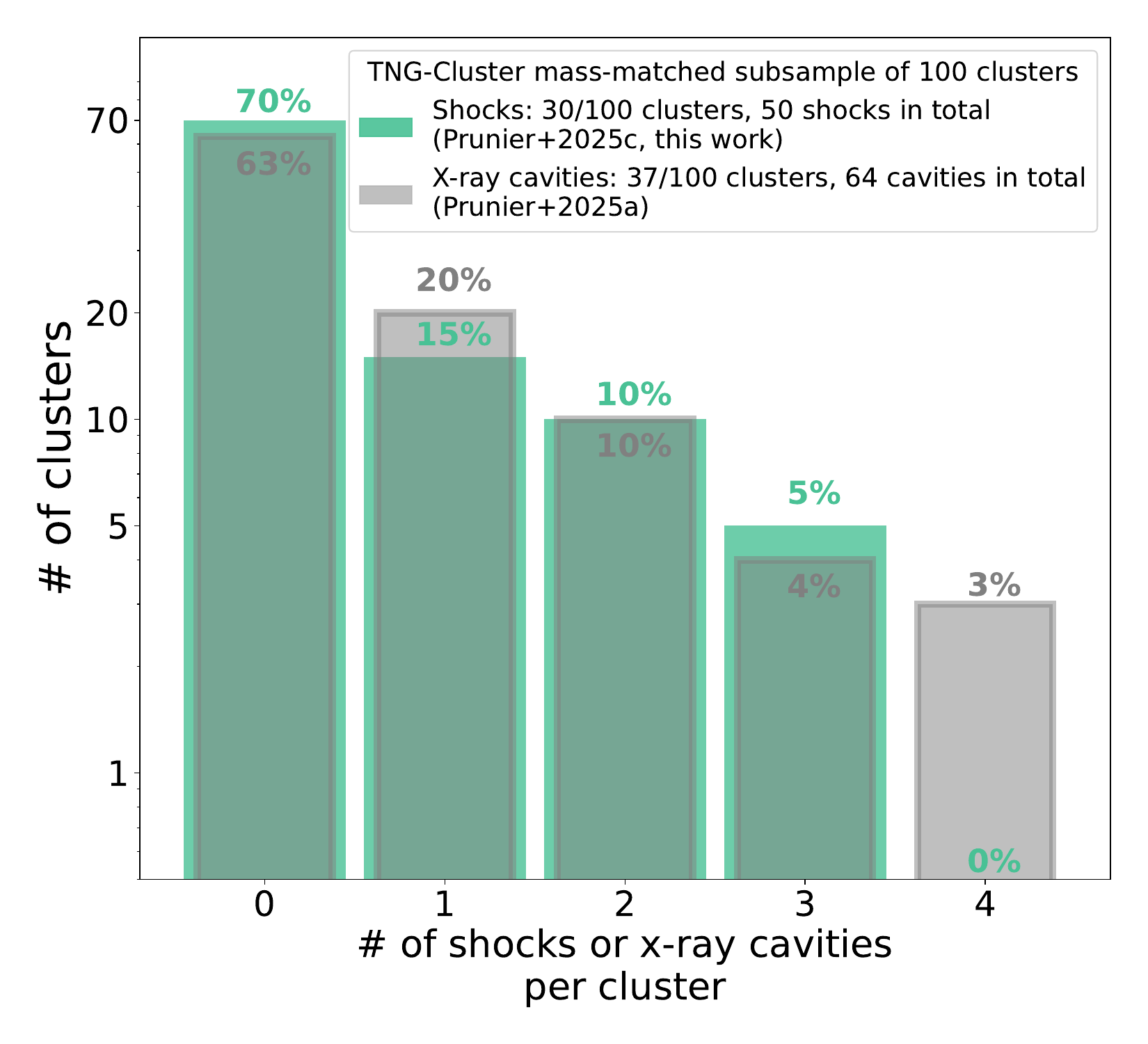}
  \end{subfigure}
  \begin{subfigure}[b]{0.35\textwidth}
   \hspace{1.2cm}
   \vspace{1.8cm}
    \includegraphics[width=1.\linewidth, height=1.\linewidth]{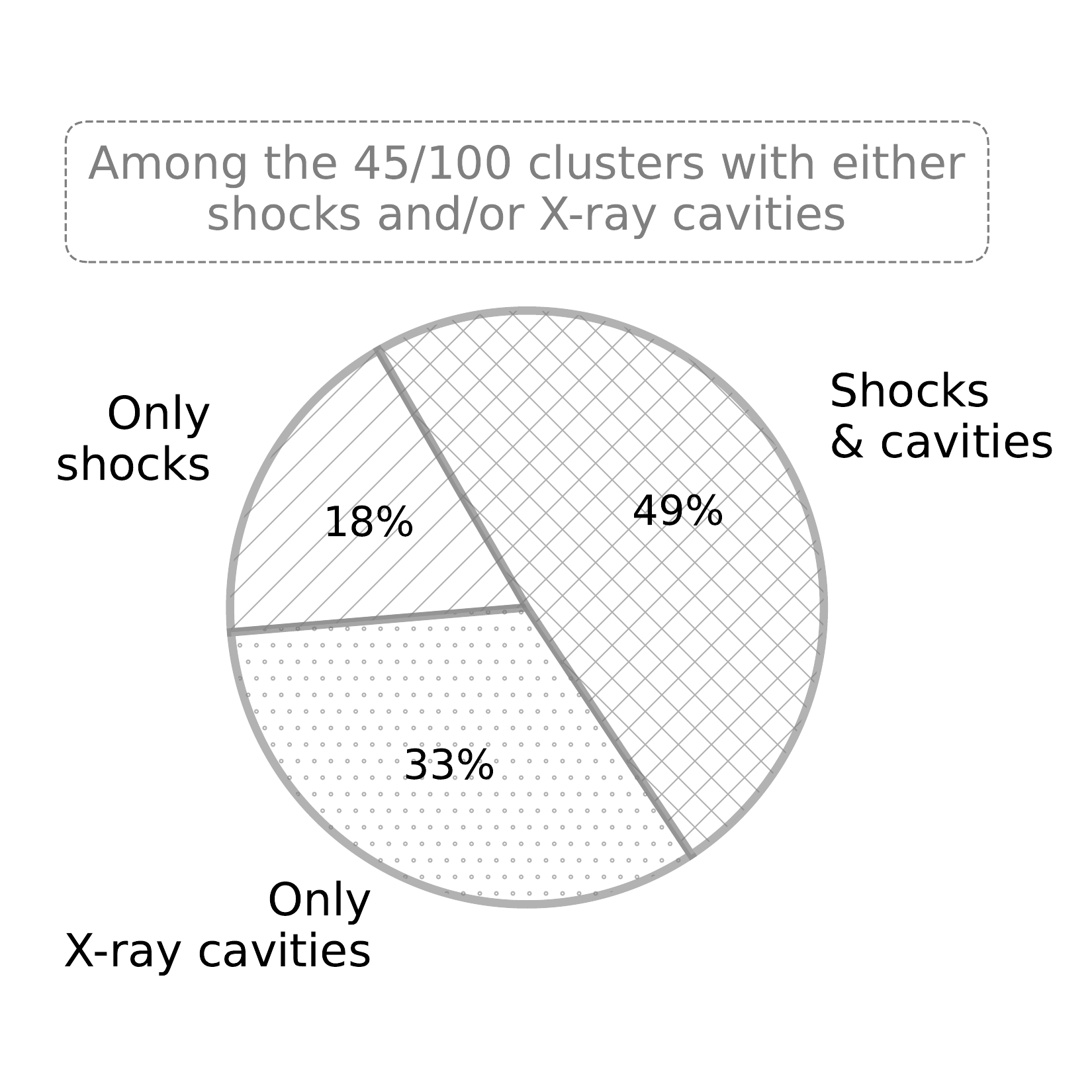}
  \end{subfigure}
   \vspace{-0.4cm}
  \caption{Demographics of clusters with and without identified shocks and X-ray cavities in the TNG-Cluster subsample of 100 clusters (at snapshot 99). \textit{Left:} Distribution of the number of shocks (mint green) or X-ray cavities (gray) per cluster, showing the percentage of clusters with none, one, two, three, or four detected features. \textit{Right:} Among the 45 clusters with at least one identified shock or cavity, the fraction hosting only shocks, only cavities, or both. Shocks are detected in 30 out of the 100 clusters in our subsample, indicating that they are {reasonably} common features in simulated clusters. While shocks frequently appear in clusters that also host X-ray cavities, about 18\% of the detected shocks are found in clusters where no cavities were previously identified in \citet{Prunier2025a}.}
  \label{fig:panel}
\end{figure*}

Shocks are not rare in the subsample {of simulated clusters from TNG-Cluster}: our systematic search in deep -- 600 ks -- mock \textit{Chandra} X-ray images reveals that one-third (30 out of 100) of clusters host one or more shock fronts, totaling 50 shocks across the sample (Figure \ref{fig:panel}). These shocks come in various morphological configurations, as highlighted in Figure~\ref{fig:panel_shocks} where we show a selection of eight clusters. In these $200\times200$ kpc images centered on the AGN, shocks appear as surface brightness discontinuities with clear enhanced edges in the filtered images. We identify early-stage shocks close to the central AGN (within $\sim$30-40 kpc), which often coincide with the edges of expanding X-ray cavities (e.g., Halo ID 15428207, Fig.~\ref{fig:panel_shocks}, Panel 3, right), as well as on larger scales, full cocoon or arc-shaped shocks (e.g., Halo ID 16207140). While some shocks are seen with or at the edge of an X-ray cavity, some shocks appear unassociated with a visible cavity (e.g, Halo ID 6834499, {Fig.~\ref{fig:panel_shocks},} Panel 5, left). This could indicate different dynamical histories, such as an outburst that generates a shock without inflating a detectable cavity, or it may be the result of projection effects that prevent the cavity from being visible in the mock image. Alternatively, the cavity may have been disrupted by surrounding gas dynamics while the shock remains visible. Multiple shocks can be present in a single system; notably, Halo ID 16207140 ({Fig.~\ref{fig:panel_shocks}}, Panel 4, left) exhibits three nested cocoon shocks.

\begin{figure*}
    \centering
        \includegraphics[width=0.46\textwidth]{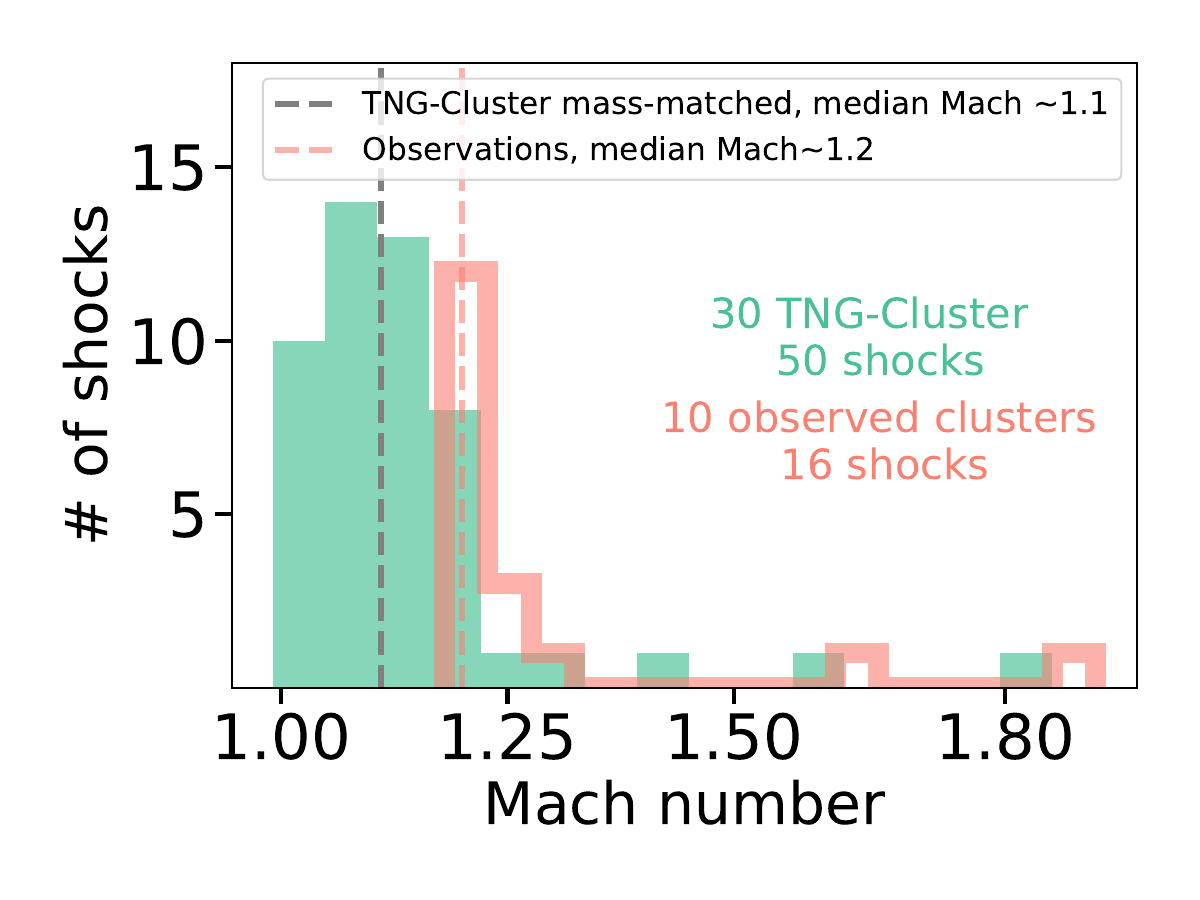}  
        \includegraphics[width=0.44\textwidth]{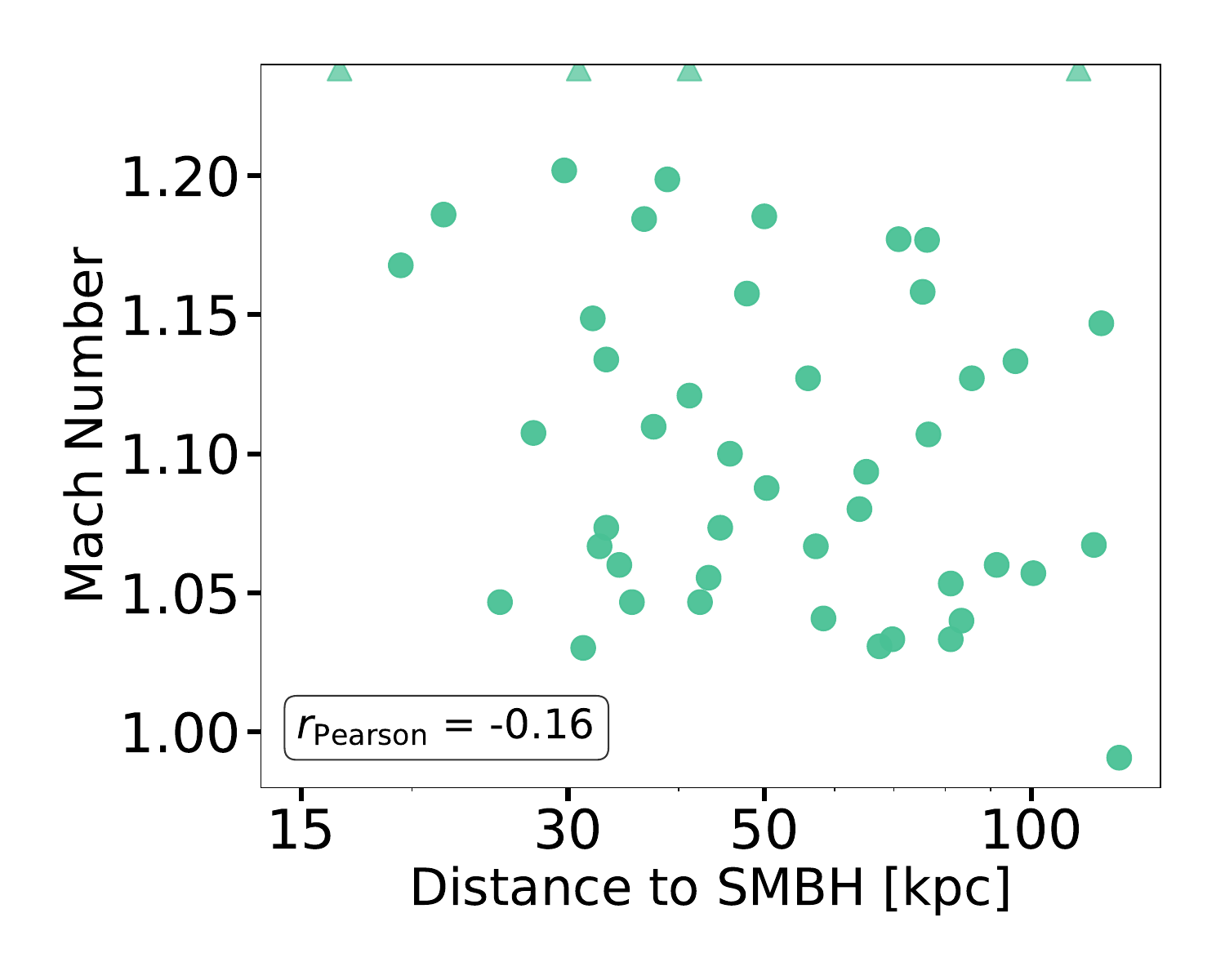}
        \vspace{-0.4cm}
        \includegraphics[width=\textwidth]{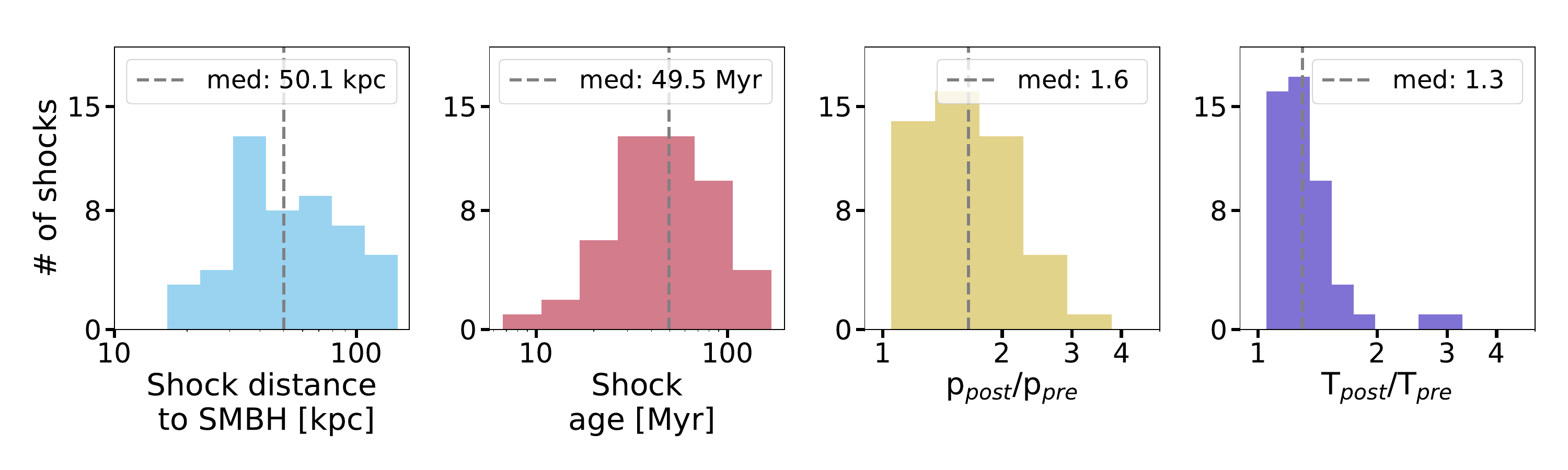}
    \caption{Shock properties in the TNG-Cluster subsample. \textit{Top Left:} Distribution of Mach numbers for the 50 shocks detected in the TNG-Cluster subsample (filled histogram) and in the observational sample (line histogram, values compiled from the literature, see Table~\ref{tab:shocks}). The simulated and observed distributions align, with comparable medians. In both cases, shocks are weak (Mach $< 2$), with 16th-84th percentile Mach 1.05–1.18, and three exceeding Mach 1.4. \textit{Top Right:} Mach number as a function of distance from the central SMBH. To emphasize the (weak) trend of decreasing Mach number with increasing distance, only the bulk of the distribution is shown (higher $>$ 1.25 Mach shocks are represented with clipped triangle markers). \textit{Bottom panels, left to right:} Distributions of shock (thermo)dynamical properties: radial distance of the shock from the SMBH, estimated shock age, pressure, and temperature jump across the shock front ({the latter two }measured from deprojected spectral fitting).}
    \label{fig:thermo_prop}
\end{figure*}

\begin{figure}
\centering
    \hspace{-0.6cm}
    \includegraphics[width=0.5\textwidth]{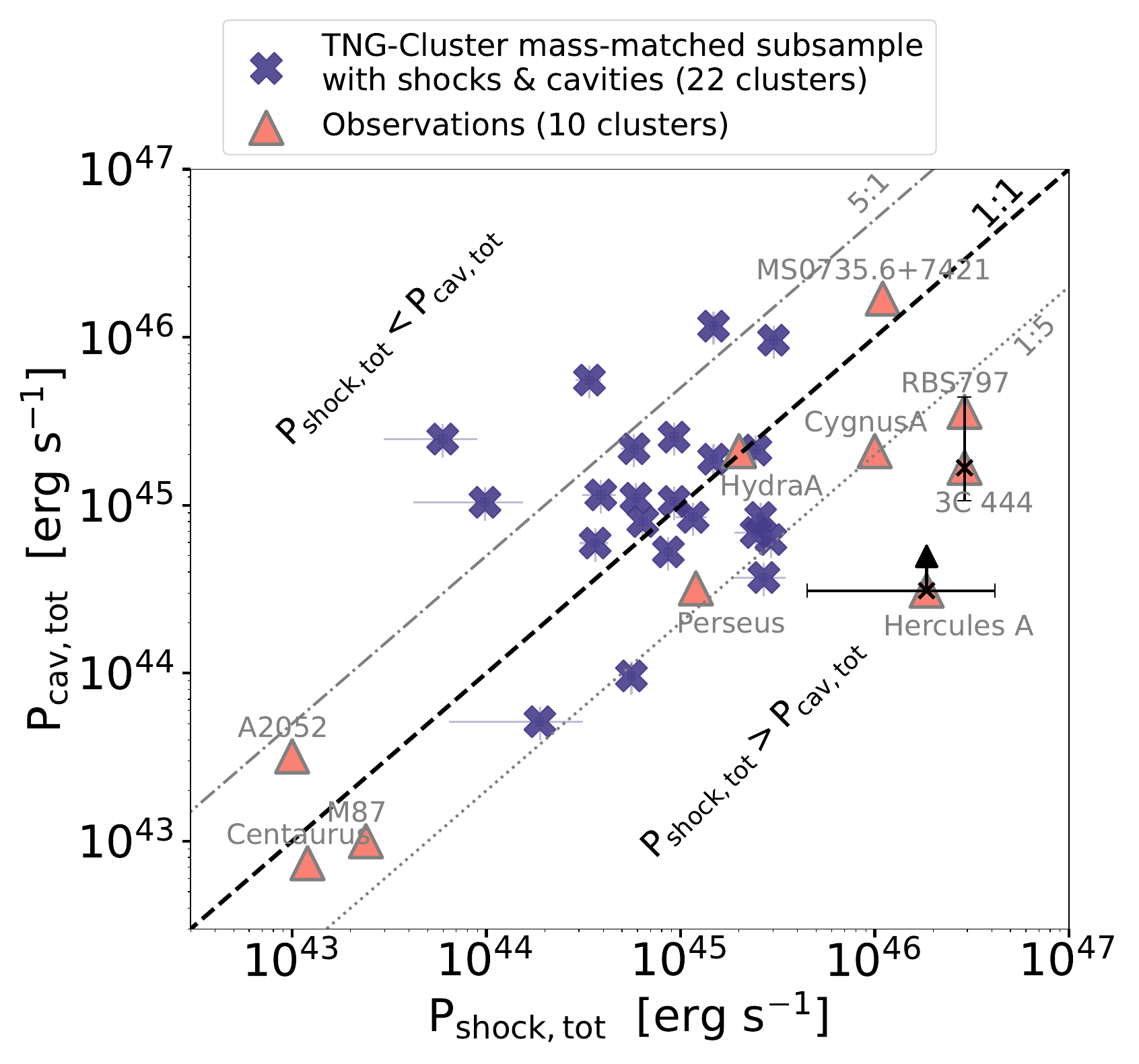}
    \caption{Total X-ray cavity power vs. total shock power for each of the 22 TNG-Cluster systems with both {X-ray cavities and shocks }(purple crosses) and for the ten observed clusters (orange triangles). Most TNG-Cluster systems have powers distributed around the $P_{\mathrm{shock,tot}} = P_{\mathrm{cav,tot}}$ line, with three clusters showing shock power ten times lower. In contrast, the majority of observed clusters have higher shock powers than cavity powers.}
    \label{fig:shock_vs_cav_power}
\end{figure}

\begin{figure*}
  \centering
    \hspace{-0.7cm}
    \includegraphics[width=0.53\textwidth]{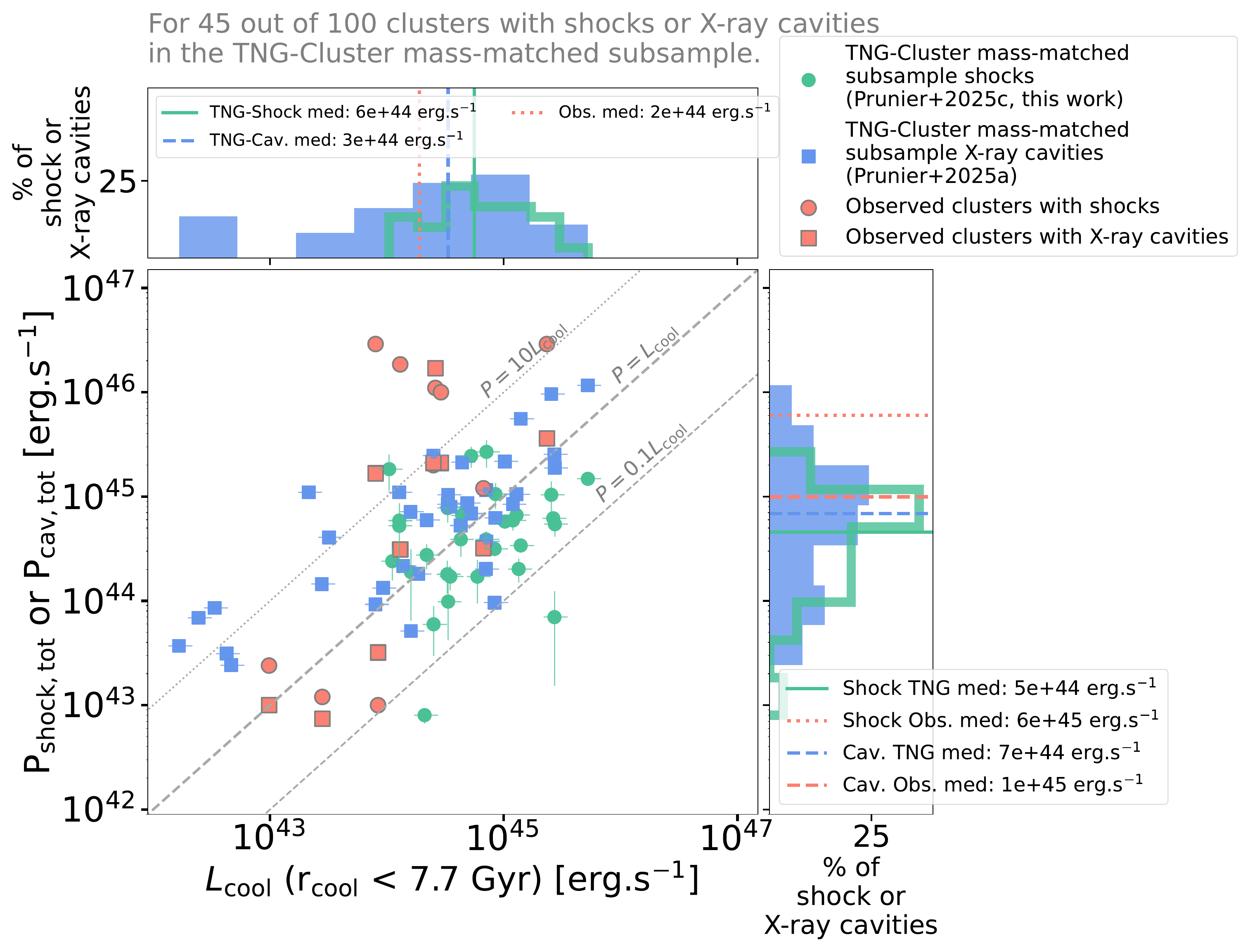}
    \hspace{-0.2cm}
    \includegraphics[width=0.51\textwidth]{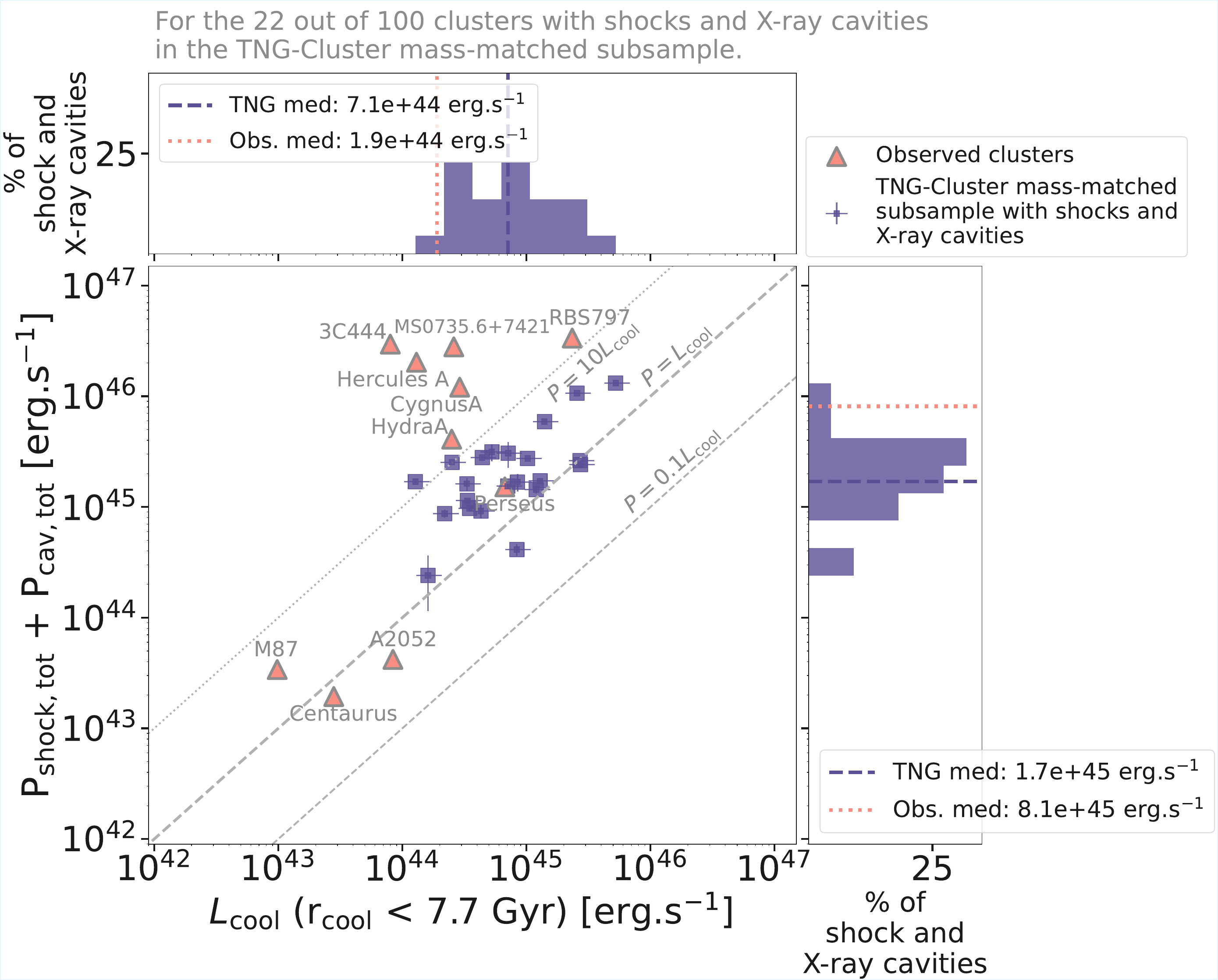}
  \caption{\textit{Left:} Total shock (P$_\text{shock,tot}$) and X-ray cavity power (P$_\text{cav,tot}$) vs. cooling luminosity (L$_\text{cool}$, for a cooling radius of 7.7 Gyr) for the 45 clusters with either identified shock or X-ray cavities (or both) in the TNG-Cluster sample (mint green circle, respectively, blue square markers) and the observational sample (values from the literature see Table~\ref{tab:shocks}). Each point represents the sum of shocks or X-ray cavities in the considered system. The dashed gray diagonal marks the 1:1 correlation between power and cooling luminosity. The side histograms display the distribution of each quantity, with the median. \textit{Right:} Total power from the summed shock(s) and X-ray cavity(ies) vs. cooling luminosity for each of the 22 clusters hosting both, and for the observational sample (triangle). In the TNG-Cluster sample, the distributions of shock and cavity power are similar and generally comparable to the cooling luminosity. For clusters with both features, the total power typically exceeds L$_\text{cool}$ by a factor of 1 to 10, which is lower than in most observed systems (10 to 100 $\times$ L$_\text{cool}$). }
  \label{fig:pcav_pshock_lcool}
\end{figure*}
\begin{figure*}
    \centering
    \includegraphics[width=0.815\textwidth]{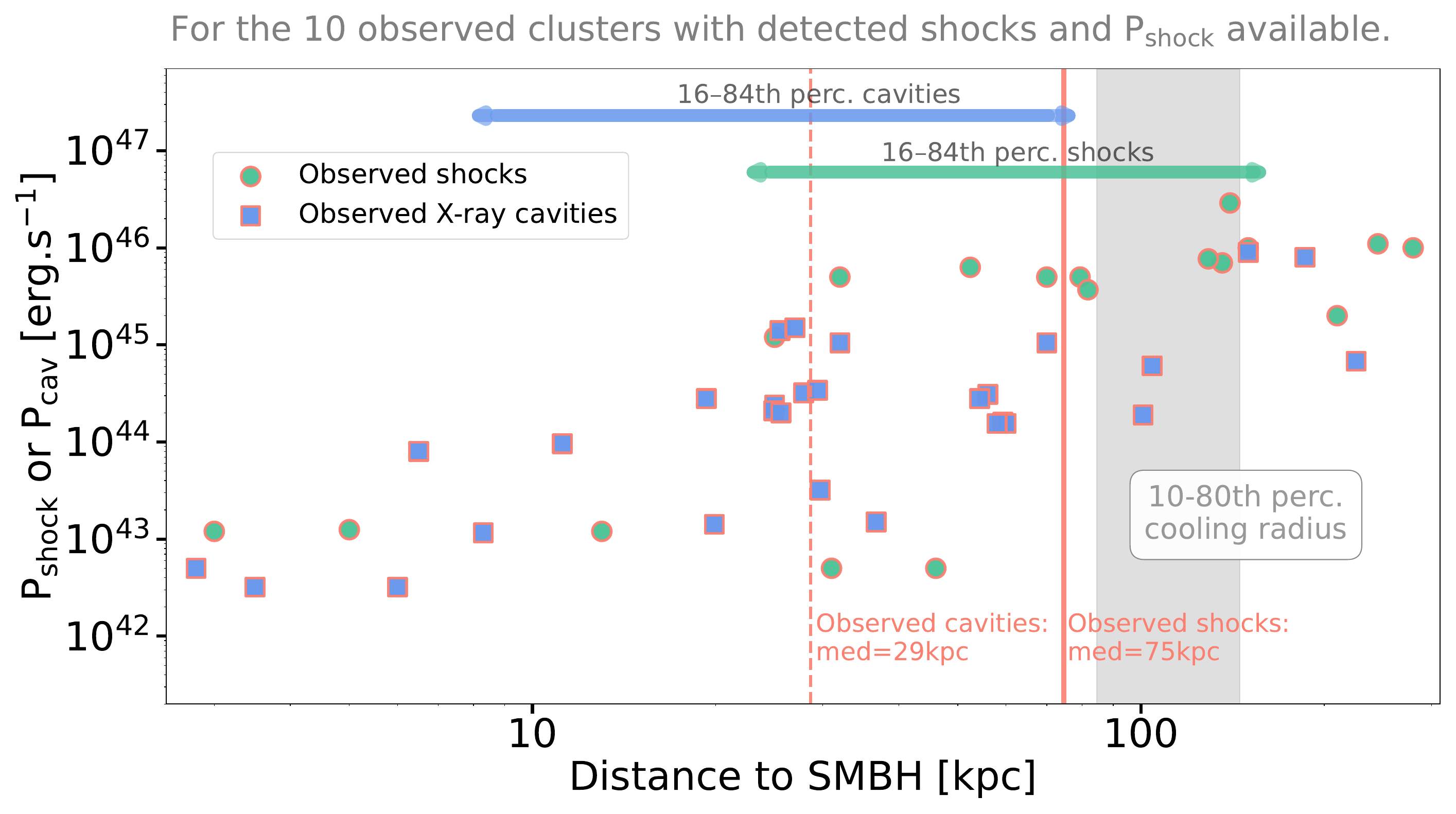}
    \includegraphics[width=0.845\textwidth]{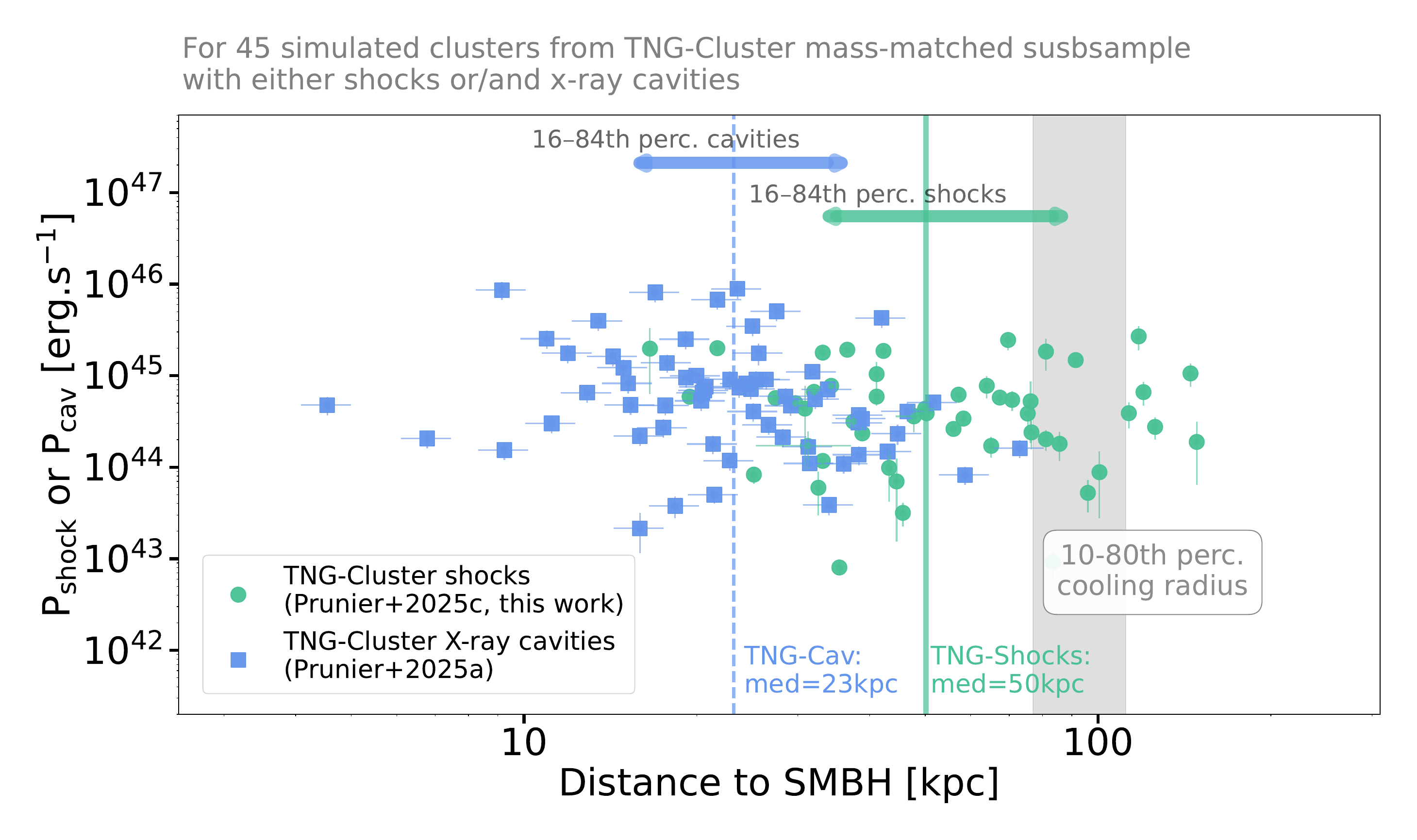}
    \vspace{-0.35cm}
    \caption{Total shock power ($P_{\mathrm{shock,tot}}$) or X-ray cavity power ($P_{\mathrm{cav,tot}}$) versus distance from the central SMBH. \textit{Top:} 9 observed clusters with detected shocks. \textit{Bottom:} 45 TNG-Cluster systems with identified shocks, cavities, or both. Shock and cavity powers are shown in lime green and blue, respectively. Median distances are indicated by dashed (cavities) and solid (shocks) lines for TNG-Cluster, and orange lines for observed clusters (for clusters where individual cavity or shock powers were unavailable in the literature, we quote the median total power values). Vertical shaded bands show the distribution of cooling radii ($t_{\mathrm{cool}} < 7.7$ Gyr) in TNG-Cluster (gray) and observations (orange). Both observations and simulations reveal a radial offset, with X-ray cavities generally found closer to the SMBH than shocks. Despite this spatial difference, the two features have comparable powers and mostly lie within the cooling radius, suggesting shocks deposit similar energy but up to larger cluster-core radii.}
    \label{fig:pcav_pshock_rad}
\end{figure*}

More quantitatively, we identify 10 TNG-Cluster systems hosting two distinct shocks, and 5 clusters with three shocks (e.g., Halo ID 11069985 or 16207140 in Fig.~\ref{fig:panel}). Observationally, most systems exhibit either a single shock (such as Perseus, MS0735.6+7421) or a pair of shocks (e.g., A~2052, M~87), with RBS~797 standing out as the only known case so far with more than two recorded shocks. We note that, in both simulations and observations, the distinction between a single elliptical (cocoon-like) shock and two separate arc-like features can be ambiguous, introducing potential uncertainties in the shock count.

Overall, we classify approximately 25\% of the detected shocks as co-spatial with an X-ray cavity -- that is, located at the boundary of an X-ray cavity -- while the remainder are identified as arc-like or cocoon-shaped shocks with no clear cavity association. {Yet, 49 \% of the detected shocks are detected in clusters that also host cavities.} Applying the same visual classification to the 16 shocks in the observed sample, we find that roughly 25–30\% also appear to be located at the edge of an X-ray cavity, or a pair of cavities. For example, the weak shock surrounding the inner X-ray cavities in Centaurus A \citep[][]{2016Sanders_CentaurusA_shock} or in M87 \citep[][]{2010Million_M87}.

When comparing TNG-Cluster halos to observed systems, we find visual analogs: Halo ID 15500142 (Fig.~\ref{fig:panel_shocks} Panel 4, right) resembles RBS~797, where central cavities are enclosed by almost closed shock front \citep[see the \textit{Chandra} image in][]{2011Cavagnolo_rbs,Doria2012,2023Ubertosi_rbs}, while the combination of multiple cavities and outer shocks in Halo ID 11069985 (Fig.~\ref{fig:panel_shocks}, Panel 1, right) mirrors the complex morphology of A~2052 \citep{2011Blanton_A2052}.

We also show in Figure~\ref{fig:panel_shocks_fit_temp_jump} the surface brightness profiles for various shocks, with a broken power-law fit to the brightness jump, as well as the corresponding temperature profile. While some shocks exhibit clear brightness discontinuities with Mach numbers exceeding 1.15 (e.g., Halo ID 12334208 or 146233683) or strong temperature jumps (e.g., Halo ID 12582631, where the temperature of the unshocked gas is roughly 1 keV lower that measured just inside the shock front (e.g., Halo ID 15500142), others are weaker, with Mach numbers between 1.0 and 1.1 and smaller temperature jumps (e.g., Halo ID 11069985). Despite their lower contrast, these weaker shocks are still statistically significant at the 2$\sigma$ level in our 600 ks mock exposures.

\paragraph*{Shocks and X-ray cavities demographic.}

Overall, both shocks and X-ray cavities seem to be relatively common in the simulated TNG-Cluster systems. In total, 45 out of the 100 clusters exhibit either shocks, cavities, or both. The full X-ray cavity population across all 352 TNG-Cluster halos has been characterized in \citet{Prunier2025a}, using mock \textit{Chandra} ACIS-I observations with 200 ks exposures. Using this previous work for our mass-matched subsample of 100 clusters, we find 37 systems hosting at least one X-ray cavity, for a total of 64 cavities. As illustrated in the pie chart of Figure \ref{fig:panel}, nearly half of the 45 clusters with X-ray features exhibit both shocks and X-ray cavities. However, a significant fraction of systems show only one of the two: 18\% host shocks without detected cavities in the mock image, while 33\% show cavities with no detected shock. This may indicate that although the two features are often related, each can occur independently in the simulation. 
It is worth noting that while shocks were identified using the deeper 600 ks mock exposures, X-ray cavities were characterized from the 200 ks images. As such, additional cavities could in principle become visible in the deeper images, which might slightly affect the statistics presented here.
In contrast, all ten observed clusters with confirmed shocks gathered in Table~\ref{tab:shocks} also host X-ray cavities, while most observed clusters in the literature \citep[e.g.,][]{2014Panagoulia_vol} show X-ray cavities without detected shocks (this asymmetry could be driven by observational biases and detection limits {and will} be discussed in Section~\ref{sec:discussion}). 

\paragraph*{Shocks are found in cool core clusters.}
We investigate the cool core status of the clusters with detected shocks and/or X-ray cavities using the classification from \citet{2024Lehle}. Specifically, clusters are defined as strong cool cores (SCCs) if their central intrinsic cooling time\footnote{For each cluster, the mass-weighted mean cooling time is computed within a volume of $0.012 \times R_{500\text{c}}$ using simulation outputs.} is below 1 Gyr, and as weak cool cores (WCCs) if the central cooling time falls between 1 and 7.7 Gyr. All clusters with detectable shocks or cavities in our sample are classified as either WCCs or SCCs (37\% WCCs, 63\% SCCs), whereas the initial 100 mass-selected clusters were more balanced, with 15\% NCCs, 52\% WCCs, and 33\% SCCs. This likely indicates a strong link between cool core conditions and the visibility or occurrence of these X-ray features. In contrast, all clusters in the observational sample are cool cores according to their respective literature.

\subsection{Mach numbers and (thermo)dynamical properties} \label{subsec:demog}

Among the 50 shocks identified in the TNG-Cluster subsample of 100 systems, Mach numbers from the surface brightness jump best fits span from 0.99 to 1.85, with a median value of 1.1, as shown in the top left histogram of Figure~\ref{fig:thermo_prop}. This distribution is broadly consistent with that of the ten observed clusters, with a median of Mach 1.2. In both simulations and observations, shocks are predominantly weak, with Mach numbers below 2 (a commonly, although arbitrary, adopted threshold to define weak shocks). However, while the distributions seem to match at the high-Mach end, no shocks below 1.2 are detected in observational data. 

The top-right panel of Figure~\ref{fig:thermo_prop} shows the Mach number as a function of the projected distance from the central SMBH. {There might be} observe a modest negative correlation for the bulk of the shock population (Mach $\sim$1–1.2, Pearson correlation index -0.16), consistent with the idea that shocks weaken as they propagate outward \citep[e.g.,][]{2005Zanni}. {Figure~\ref{fig:thermo_prop} combines shocks from different AGN outbursts with varying energy outputs, which naturally produces a wide spread in Mach number at any distance from the central AGN. By contrast, following a single shock at different times, as shown in Figure \ref{fig:dicsu_mach}, illustrates the theoretically expected decrease in Mach number with time as the shock propagates into the ICM.} Observed shocks (not shown {in  Fig.~\ref{fig:thermo_prop} for clarity}) occupy a similar region in Mach–distance space but do not exhibit a clear anti-correlation. 

In terms of projected distance from the central SMBH, shocks in the TNG-Cluster sample are {typically found at distances of a few tens of kpc up to $\sim$150 kpc, with the most distant shock at 149 kpc,} and have estimated ages of a few tens of Myr, with a median age of $\sim$46 Myr. Observed shocks occupy a similar distance range, though three shocks are found at larger distances ($>$150 kpc).

Across the sample, we find temperature and pressure jumps before (lower temperature and lower pressure) versus after (higher temperature due to shock-heated gas, higher pressure due to gas compression) the shock passage with typical values $\sim$1.3 and $\sim$1.6, respectively (median values). These numbers broadly match the shock literature \citep{2007Forman,Blanton2009_2052,2009Simionescu_HydraA,Croston2011_3C444,snios2018cocoon-a43,2023Ubertosi_rbs,Ubertosi2025_HerculesA}. However, since such measurements are often not available for observed shocks, we do not include them {as} observational comparison {in the} plots.

Although not shown in the panel, there exists a positive correlation between the Mach number computed from the surface brightness jumps and the temperature and pressure jumps. We find that higher jumps are associated with stronger shocks, a result that is expected from the Hugoniot-Rankine conditions \citep[][]{1967RankineHug} and also found in the literature \citep[e.g.,][]{2009Simionescu_HydraA,snios2018cocoon-a43}. This may reflect the local heating and compression caused by the shock, as the jumps were measured in narrow regions immediately upstream and downstream of the shock (Method \ref{subsec:meth_shock_prop}).

\subsection{Weak shock versus X-ray cavity powers} \label{subsec:shock_power_vs_cav}

To assess the relative importance of shock vs. X-ray cavity energetics, we compare their respective inferred powers in the 22 TNG-Cluster systems of our subsample that host both features. Shock power estimates the rate of energy dissipation in the intracluster gas, calculated from the pressure jump across each shock \citep[see Method and, e.g.,][]{1959Landau}. In contrast, cavity power reflects the mechanical work required to inflate each X-ray cavity against the surrounding gas pressure \citep[see Method and, e.g.,][]{hlavacek-larrondo_agn_2022}. Both quantities are measured using the observational techniques\footnote{In observational studies, the gas within X-ray cavities is typically assumed to be a relativistic plasma (adiabatic index $\gamma = 4/3$), whereas in TNG, we adopt $\gamma = 5/3$, consistent with an ideal monatomic gas. We verified that using $\gamma = 4/3$ for TNG would increase the measured cavity powers by less than 0.2 dex.} on the \textit{Chandra} mocks {of TNG-Cluster halos}.

We find that both shock and cavity powers span the range $10^{44}$–$10^{46}$ erg s$^{-1}$ in these TNG halos, which {are} broadly consistent with typical X-ray cavity and shock powers reported in the literature \citep[e.g.,][]{2014Panagoulia_vol,2023Ubertosi_rbs}. In Figure~\ref{fig:shock_vs_cav_power}, we plot the total shock power versus the total cavity power (i.e., summing all shocks, respectively cavities, per system) within each of the 22 TNG-Cluster systems that host both features. The distribution broadly scatters around the 1:1 line ($P_{\mathrm{shock,tot}} = P_{\mathrm{cav,tot}}$), suggesting that, on average, shocks and cavities contribute comparably to the energy budget in the simulation. However, the scatter is large --- up to $\pm$1 dex --- indicating significant variation in how energy is distributed in individual systems. 

We also include the ten observed clusters with confirmed shocks in the same plot (triangle markers). Their distribution is skewed toward $P_{\mathrm{shock,tot}} > P_{\mathrm{cav,tot}}$, with 7 out of 9 systems showing higher shock than cavity powers, although this difference is mostly within 1 dex. The most extreme case is 3C~444, where the shock power exceeds the cavity power by a factor of $\sim$100. However, for some systems, such as 3C~444 and Hercules A, the estimated cavity power is {likely} affected by extended non-thermal X-ray emission arising from inverse Compton (IC) scattering of CMB photons by relativistic electrons in the radio lobes, which can partially obscure the cavities. In 3C~444, IC emission introduces large uncertainties in the cavity power estimates \citep{Croston2011_3C444}, while in Hercules A, the inferred cavity power should be considered an upper limit \citep[][]{Ubertosi2025_HerculesA}. In cases with large uncertainties in either cavity or shock power, we indicate them in Figure~\ref{fig:shock_vs_cav_power} with error bars and arrows.

\subsection{Comparing weak shock and X-ray cavity powers to cluster cooling luminosities} \label{subsec:power_vs_cool}

We compare the powers associated with shocks and X-ray cavities in the TNG-Cluster subsample to the cooling luminosity within the cluster core. The cooling region (r$_\text{cool}$) is defined as the radius within which the gas cooling time is shorter than 7.7 Gyr, consistent with the one adopted in the literature for the ten observed clusters listed in Table~\ref{tab:shocks}. This region broadly corresponds to where feedback is expected to regulate the thermodynamic state of the intracluster gas, {and ranges between 57-124 kpc for the 100 studied TNG-Cluster systems \citep[see also][]{Prunier2025b}}. The vast majority of observational studies focus only on X-ray cavities, finding that their energy output often balances ICM cooling, typically yielding a near 1:1 correlation between cavity power and cooling luminosity \citep[e.g.,][]{2008Birzan}.\\

In Figure \ref{fig:pcav_pshock_lcool} left panel, we show the total shock and/or cavity powers per cluster for the 45 TNG-Cluster systems with either shocks or cavities, against the cooling luminosity ({note that all quantities are} estimated using observational methods; see section \ref{subsec:meth_shock_prop}, \ref{subsec:cav}, \ref{subsec:met_cool_lum}), and the same quantities for the ten observed clusters (triangles). Therefore, for each system, up to two data points are shown: one corresponding to the total shock power ($P_{\mathrm{shock,tot}}$), obtained by summing the contributions from all detected shocks (if present), and one to the total cavity power ($P_{\mathrm{cav,tot}}$), obtained by summing over all detected cavities (if present), each plotted as a function of the same characteristic cooling luminosity of the cluster ($L_{\mathrm{cool}}$). 

Focusing first on the P$_\text{shock,tot}$ vs. L$_\text{cool}$ in the left panel of  Fig.~\ref{fig:pcav_pshock_lcool}: in TNG-Cluster (mint green circles), the shocks broadly scatter around the 1:1 line (central dashed), with a median power of P$_\text{shock, tot} \sim L_{\rm cool} \sim (5$–$6)\times10^{44}$ erg/s across the 30 clusters hosting shocks. However, in the few observed clusters with detected shocks, most have $P_{\mathrm{shock, tot}} / L_{\mathrm{cool}} \geq 1$, indicating {that} mechanical feedback largely balances cooling. The observed medians (orange) show shock power about 30 times higher than the cooling luminosity.

Focusing now on P$_\text{cav,tot}$ vs. L$_\text{cool}$ in the left panel of  Fig.~\ref{fig:pcav_pshock_lcool}: in the TNG-Cluster subsample (blue squares), the 64 X-ray cavities roughly follow the 1:1 correlation with $L_{\mathrm{cool}}$, with median powers comparable to cooling luminosities: $7 \times 10^{44}$ erg/s versus $3 \times 10^{44}$ erg/s, and overlapping the parameter space of shocks. Overall, cavity powers and cooling luminosities align well with observations, showing similar trends and medians \citep[consistent with the Fig.9 of][for a different TNG-Cluster subsample, in a ``apple-to-apple'' comparison with observations]{Prunier2025b}.

As shown in the previous section and Fig.~\ref{fig:shock_vs_cav_power}, observed shocks are often more powerful than X-ray cavities; the median shock power in the observed sample exceeds the median cavity power by $\sim$1.5 dex. In contrast, TNG-Cluster shocks are, on average, as powerful as the X-ray cavities, and align with the cooling luminosities.\\

Focusing now on the right panel of Figure \ref{fig:pcav_pshock_lcool}, we isolate the 22 clusters having both detected shocks and cavities in our TNG-Cluster subsample. We compare the total mechanical power ($P_{\text{tot}}$ = P$_\text{shock,tot}$ + P$_\text{cav,tot}$) to L$_\text{cool}$ for each. Most observed systems (6 out of 9) occupy the upper region of the P$_\text{tot}$/L$_\text{cool}$ plot, with ratios ranging from 14-16 (RBS~797, Hydra A) to 370 (3C~444). In contrast, only one of the 22 TNG-Cluster systems exceeds $L_{\text{cool}}$ by more than an order of magnitude. While most TNG-Cluster systems lie above the 1:1 line, indicating that mechanical feedback is sufficient to offset or exceed radiative cooling, their $P_{\text{tot}}/L_{\text{cool}}$ ratios remain confined within a factor of $\sim$1–10, lacking the most extreme outliers seen in the observed sample.

\subsection{A dominant feedback channel at different radii} \label{subsec:domi_heat_channel}

In both the TNG-Cluster simulation subsample and observed clusters, the powers of shocks and X-ray cavities are comparable; however, their spatial distributions relative to the central SMBH differ significantly. X-ray cavities are generally found closer to the SMBH, while shocks are located at greater distances, as {we quantify} in Figure \ref{fig:pcav_pshock_rad}. This is reflected in the median distances, which are 23 kpc and 50 kpc for cavities and shocks, respectively, in the simulation, and 29 kpc and 75 kpc in the observational sample.

The observational dataset, being smaller (16 vs. 50 shocks and 30 vs. 64 cavities in the simulation), exhibits larger scatter and broader 16th–84th percentile ranges in the distance distributions (blue and green arrows,  Fig.~\ref{fig:pcav_pshock_rad}). In contrast, the simulations show tighter distributions (lower panel). Despite these differences in sample size and variance, both datasets consistently display a clear {distance} offset between shocks and cavities relative to the central AGN. To our knowledge, this spatial distinction is reported here for the first time.

To first order, this result gives us insight into the different propagation characteristics of these two features: shocks move outward at roughly the local sound speed ($\mathcal{M} \sim$ 1.1–1.2), while cavities, which are on average found at radii more than a factor of two smaller than the shocks, rise more slowly, at subsonic speeds. This interpretation is supported by one observational result in A~2052, where \citet{2018Balmaverde} directly measured the velocity of line-emitting gas at the edge of the northern cavity with MUSE, and found that the cavity {is} likely expanding subsonically.

Furthermore, most X-ray cavities (all of them in the TNG-Cluster subsample) and most shocks lie within the cooling radius (where t$_\text{cool} < $ 7.7 Gyr). It suggests that shock fronts, while propagating beyond the central tens of kpcs, typically do not reach beyond the cooling radius before becoming undetectable (but see outer shocks in MS0735.6+7421, Hydra~A, and RBS~797), either due to broadening into sound waves or insufficient X-ray photon statistics \citep[see also][]{2013Nulsen_sh_sound_heat}. This could constrain the effective range of shock heating and may imply that cavities deposit energy closer to the cluster core, while shocks transport energy from the center up to larger radii.

\section{Discussion and implications}\label{sec:discussion}

The handful of observed clusters with confirmed shock fronts likely represent just the tip of the iceberg for AGN-driven shocks in cluster cores. The lack of statistics and demographic information leaves many key questions open, including how prevalent these shocks are and what role they play in heating cluster cores. 

Our analysis of 100 deep mock \textit{Chandra} observations {of} the TNG-Cluster simulation predicts that shocks are far from rare, as we detect them in approximately one-third of the clusters. We also find that they contribute significantly to the feedback, with the mechanical power carried by shocks often comparable to that inferred from cavities. These findings, along with the shock properties, broadly agree with those from the ten observed clusters with confirmed shocks. However, TNG-Cluster shocks exhibit slightly lower Mach numbers (median $\mathcal{M} \sim$ 1.1, versus 1.2 for observed shocks), while observational data show shock powers that are generally higher than cavity powers and exceed the cooling luminosity (for instance, 5 out of 16 observed shocks have powers more than 10 times above L$_{\mathrm{cool}}$, a case not found in the TNG shocks). Most strikingly, we find that shocks are preferentially located at larger distances from the central AGN than cavities, a spatial separation that is also present in observations. This suggests a complementary role for shocks and cavities in AGN feedback, where shocks deliver energy more isotropically and at greater distances, working in synergy with cavities to regulate cooling in cluster cores.

Whether the shocks identified in TNG-Cluster fully represent the shock population that might exist in real clusters remains an open question. The limited number of observed clusters with confirmed shocks, combined with unknown selection biases, complicates direct comparisons. It is also unclear if these ten observed systems capture the full diversity of shock properties, making it difficult to assess whether discrepancies between simulations and observations are statistically significant. Moreover, while observed shocks are often linked to AGN-driven radio emission, shocks in the simulation may arise from a more complex mix of processes, including merger-driven phenomena {\citep[see][for radio relics in TNG-Cluster]{2024Lee}}.

However, if TNG-Cluster shocks do reflect reality to some extent, our findings, combined with existing observational evidence, support an emerging view that shocks play a key role in cool core clusters.\\

To {further} address these points, we first discuss how representative the TNG-Cluster shock population might be in {Section} \ref{sec:discussion_represent}, commenting on the origin of the identified shocks, the realism of the simulated clusters and of the TNG feedback model, and on the potential selection biases in observations. We then explore the broader implications of our findings for cluster core heating in {Section} \ref{sec:disc_cycle}, and last, prospects for future observations with next-generation X-ray facilities.

\subsection{Are TNG-Cluster shocks representative of the real population?}\label{sec:discussion_represent}

\paragraph*{On the nature of the detected shocks.}

X-ray shocks in galaxy clusters can arise from two main phenomena: mergers or AGN activity. While our detection pipeline, as well as those employed in the literature, can not explicitly distinguish between these origins, certain properties can help identify the likely nature of the shocks. Here, we assess whether the 50 shocks identified in this work may include a portion of merger-driven shocks.

From the literature on shocks, merger-driven shocks are typically found in dynamically disturbed, non-relaxed clusters, often associated with sloshing patterns \citep[e.g., A2142][]{2000Markevitch_A2142,2013Rossetti}, and tend to have Mach numbers greater than 2, as observed in clusters like A520 \citep{2005Markevitch_A520}, and the Bullet Cluster \citep{2007Milosavljevic_Bullet}. 
In contrast, the 16 observed shocks gathered in this work (Table \ref{tab:shocks}) all reside in the centers of relaxed, cool core clusters. In the TNG-Cluster subsample, all detected shocks occur exclusively in cool core clusters, with over 60 \% residing in strong cool cores, indicative of a relaxed dynamical state. {Importantly}, visual inspection of their mock \textit{Chandra} images reveals no evidence of minor mergers associated with the shock fronts

In terms of spatial location and morphologies, merger shocks are generally located above 500 kpc from the cluster center, and morphologically appear as two caps but rarely as a closed cocoon centered on the AGN \citep[e.g.,][]{2025Lee}. In the TNG-Cluster subsample, all the detected shocks are located close to the cluster center ($<$150 kpc), and the majority appear as arced surface brightness jumps with centers offset by less than 3 kpc from the central SMBH, as determined by elliptical fits to their shapes. About 40\% of the detected shocks form a nearly closed arc or cocoon structure (panda \texttt{CIAO} region with closure angle >200°), surrounding either the SMBH or an X-ray cavity, consistent with being inflated by SMBH activity.

We conclude that contamination from merger shocks is negligible among the 50 shocks detected in TNG-Cluster, allowing us to confidently attribute these shocks to AGN feedback.

\paragraph*{Fidelity of TNG-Cluster gas properties and AGN feedback modeling.}

We identify two differences between the shock populations in the TNG-Cluster subsample and the observed sample. 

First, we detect a population of nearly sonic, weak shocks with Mach numbers below 1.1, which are absent from the current observational literature. This discrepancy may arise from (i) observational challenges: such weak shocks are inherently difficult to detect and may be overlooked due to selection biases that favor systems with stronger, more prominent features; or (ii) physical differences, whereby shock waves in real clusters dissipate more rapidly than in the simulations. Nearly sonic shocks are likely to be present in real clusters, but their detection is hindered both by their subtle observational signatures and by the lack of dedicated searches specifically targeting them.

Second, we do not recover the most powerful shocks reported in some observations, as shown in Figures~\ref{fig:shock_vs_cav_power} and~\ref{fig:pcav_pshock_lcool}. This discrepancy could be partly due to selection effects, {as} observational studies tend to focus on clusters where cavities are present or shocks are suspected, but also to observational constraints (see next paragraphs). However, this lack of high-power shocks in the TNG-Cluster subsample cannot be fully attributed to these biases. In our mock \textit{Chandra} observations, constructed at fixed redshift $z=0.1$ and with comparable exposure times, none of the detected shocks reach the power levels observed in the most extreme real systems. One possible interpretation is that the TNG model underproduces these extreme shocks, potentially due to limitations in the AGN feedback implementation or cluster thermodynamic properties.\\

TNG-Cluster utilizes the unchanged TNG model, which is a robust and well-validated physical model for galaxy formation and evolution, as described in \citet{2017Rainer} and \citet{2018Pillepich}. The global thermodynamic and kinematic properties of the intracluster gas in the TNG-Cluster halos have been shown to align reasonably well with observations \citep{2024Nelson,2024Ayromlou,2024Lee,2024Truong,2024Lehle,2024Rohr}. Profiles of temperature, density, entropy, and pressure match observed trends, as do cool core fractions \citep{2024Lehle}, multiphase gas kinematics \citep{2024Ayromlou}, and turbulence levels \citep{2024Truong}. Furthermore, in \citet{Prunier2025b}, we show, {via} an ``apple-to-apple'' comparison, with a carefully matched observational sample that the temperature, density, and cooling properties of the gas in the core, and at the X-ray cavities location, are consistent.

The shocks and X-ray cavities detected in the TNG-Cluster subsample naturally result from the SMBH subgrid feedback model implemented in TNG. At low redshift, all cluster-central SMBHs operate in the kinetic mode, featuring episodic, momentum-driven, re-orienting outflows spaced by tens of Myr. As shown in \citet{Prunier2025b}, this model produces X-ray cavities broadly consistent with observations, and caused by SMBH kinetic power release in the range of $10^{42} - 10^{45}$ erg s$^{-1}$ \citep{Prunier2025a}, suggesting that SMBH feedback is not drastically over- or underpowered compared to observational constraints. However, in \citet{Prunier2025b} we reported a lack of small ($<$10 kpc) cavities, which could, among other possibilities, result from too energetic SMBH outbursts. In contrast, the scarcity of high-power shocks found here may indicate that shock-generating feedback is slightly underpowered. Still, disentangling the role of the feedback model from the thermodynamic and kinematic state of the gas remains challenging.

The morphology and energetics of simulated shocks are shaped by both the SMBH feedback model and the ICM's thermodynamic state. Despite these model-specific effects, past studies show that TNG-Cluster reproduces observed X-ray cavities and gas properties well, supporting the realism of its galaxy formation model at all scales, and, by extension, of the detected AGN-driven shocks.

\paragraph*{Is the detection rate higher for high $P_{\rm shock}$?}

Although none of the detected TNG-Cluster shocks reach the extreme powers ($>10^{46}$erg s$^{-1}$) seen in some observed systems, this discrepancy may be amplified by an observational bias that favors the detection of more powerful shocks. To test this, we select five clusters (out of 22 with both shocks and cavities) spanning a range of Mach numbers (1.05-1.4), shock positions ($\sim$ 20–150 kpc), and powers ($3.3 \times 10^{44}$ to $2.2 \times 10^{45}$erg s$^{-1}$). We then generate mock \textit{Chandra} images with a shorter 100 ks exposure to assess which shocks remain detectable.

Out of the ten shocks in this subsample, those with $P_{\rm shock} < 10^{45}$erg s$^{-1}$ (three in total, all with low Mach numbers, $<1.15$) are no longer detectable in surface brightness profiles at 100 ks. Interestingly, two shocks with similar powers ($\sim 10^{45}$erg s$^{-1}$) show different detectability: the one at $\sim 40$ kpc remains visible, while the one at $\sim 120$ kpc does not. This highlights that shocks at larger distances from the AGN, where surface brightness is lower, are also more easily missed.

In contrast, all ten cavities in the same systems remain detectable at 100 ks. These cavities span a power range of $10^{44}$ to $5 \times 10^{45}$erg s$^{-1}$ and are all located within 50 kpc of the central AGN.

This test on 10 shocks suggests a detection bias toward higher-power shocks {in the real Universe}, implying that the observed shock population likely represents the upper end of the intrinsic distribution. 

Additionally, although we do not test the redshift dependence here, several of the most powerful observed shocks are found in higher-redshift systems, e.g., MS~0735+74 ($z=0.25$), RBS~797 ($z=0.35$), 3C~444, and Hercules~A ($z=0.15$). At such redshifts, observational sensitivity is naturally biased toward detecting only the most prominent (and energetic) shock fronts.\\

{Projection effects could also play a role in detectability. AGN-driven shocks are expected to propagate nearly isotropically in a spherical shell around the AGN, as predicted by point explosion models and supported by systems such as RBS~797, M87, and Cygnus~A. This geometry implies that projection should not significantly affect shock detectability, although it may influence shock power estimates. To test this, we examined three TNG-Cluster systems with multiple shocks along four different projections each. In all cases, shocks were recovered at consistent radii and Mach numbers, but the apparent arc length varied with orientation. Since $P_{\rm shock} \propto V_{\rm shock}$, this variation can bias the inferred power, so our quoted values should be regarded as lower limits. Projection effects impact both observations and simulations, and may contribute to the scatter in Figure~\ref{fig:pcav_pshock_lcool}. Still, they cannot by themselves explain the lack of counterparts to the most powerful observed shocks in the TNG-Cluster matched subsample.}

\paragraph*{Sample selection considerations.}

The observed sample of clusters with shocks is likely subject to significant selection biases. \textit{Chandra} and/or \textit{XMM-Newton} deep observations often target particularly cool core {systems} showing extreme feedback, such as those with prominent or numerous X-ray cavities like RBS~797 or Hydra~A. In many cases, these clusters were first identified as hosting X-ray cavities \citep[e.g.,][]{2004Birzan}, and were later followed up with deeper X-ray observations, during which shocks were subsequently detected. Consequently, shock detection, which requires long integration times, is biased toward clusters already known to exhibit strong AGN feedback signatures. A comprehensive analysis of the \textit{Chandra} and \textit{XMM-Newton} archive could better quantify this selection bias, but such an investigation is beyond the scope of this work. Conversely, the TNG-Cluster subsample in this study was selected solely on halo mass within R$_\text{500c}$ to broadly match the range of the ten observed clusters with confirmed shocks. This relatively unbiased selection approach enables a more general comparison than the \textit{Chandra} observed sample, with an unknown and potentially complex selection function. 

\begin{figure}
\centering
    \hspace{-0.8cm}
    \includegraphics[width=0.5\textwidth]{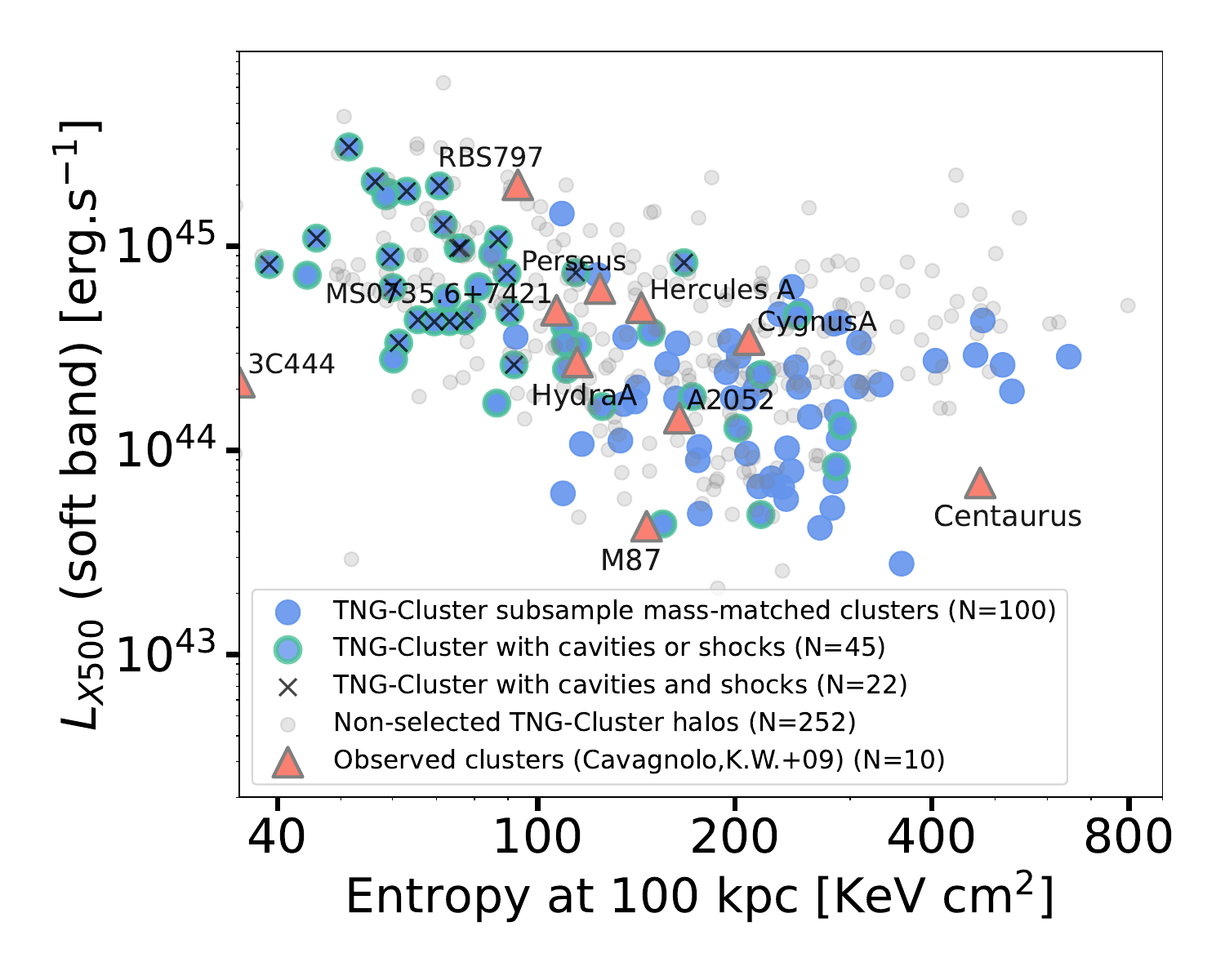}
    \caption{ Soft-band X-ray luminosity within R$_{500c}$ versus core entropy ($K = k_B T n_e^{-2/3}$, measured at 100 kpc from the center) for the TNG-Cluster systems (emission-weighted) and ten observed clusters \citep[][]{2009Cavagnolo} with detected shocks and X-ray cavities. The entropy value for 3C~444 is not available in the literature and is clipped at the x-axis limit. Clusters with both shocks and cavities in TNG-Cluster and in the observed sample have lower core entropy, with 16th–84th percentile of 57–90 keV cm$^2$ and 97–160 keV cm$^2$, respectively.}
    \label{fig:disc_sel_bias}
\end{figure}

For instance, in this work, while shocks are found across the full M$_\text{500c}$ range, with no clear dependence on cluster mass, they are more common in cool core systems and, as shown in Fig.~\ref{fig:disc_sel_bias}, occur in clusters that are thermodynamically distinct in entropy and X-ray luminosity. We compare entropy measurements at 100 kpc from \citet{2009Cavagnolo} (triangles) values for the TNG-Cluster systems\footnote{Emission-weighted entropy extracted in a shell with thickness 4 kpc. For a subset of clusters, we verified that the emission-weighted entropies are consistent with spectrally fitted values.}. TNG-Cluster halos with shocks or cavities (green edges) lie at lower entropy and higher luminosity than the rest of the sample (blue). This trend becomes even more pronounced when isolating clusters that host both shocks and cavities (marked with black crosses), which are at the tail of the low-entropy distribution (38–160 keV cm$^2$, median 71), even below the observational sample (93–210 keV cm$^2$, median 125). This supports a feedback loop in which lower-entropy clusters, cooling more rapidly, require more frequent or powerful AGN outbursts to maintain thermal balance.

This suggests that, in the TNG-Cluster unbiased subsample of 100 clusters, shocks preferentially arise in low-entropy, high-luminosity systems, implying that, despite its limited size, the current observational sample may be broadly representative of the underlying shock-hosting cluster population.

\paragraph*{Final thoughts on representativeness.}

By performing a blind search for AGN-driven shocks in a large, mass-selected sample of clusters from the cosmological TNG-Cluster simulation, with deep, 600 ks, mock \textit{Chandra} images at fixed redshift ($z=0.1$), we minimize many of the biases that affect current observational studies. Both in the simulation and in the limited sample of observed clusters, shocks and cavities preferentially occur in cool core systems with low entropy and high X-ray luminosity. Shock morphologies, properties, and scaling relations broadly align with observations, except for the high-power end, though this may reflect an observational bias favoring strong, centrally located shocks. Despite the small size and potential selection biases of the observational sample, this overlap suggests that TNG systems are rather representative of the typical conditions under which shocks arise in galaxy clusters.\\

{Interestingly, for the ten observed clusters, the ratio of shock power to cooling luminosity depends on galaxy radio class \citep{2023Ubertosi_rbs}. FR~I, with less collimated, lower-power jets that dissipate energy near the cluster center, typically show $P_{\rm tot}/L_{\rm cool} \lesssim$10–20, while FR~II or hybrid systems (e.g., 3C~444, Hercules~A, Cygnus~A) can exceed $P_{\rm shock} = 100 L_{\rm cool}$. The $P_{\rm shock}/L_{\rm cool}$ ratios in the TNG-Cluster subsample align more with FR~I-like systems. However, this should be interpreted with caution, as the TNG kinetic feedback model drives weak shocks and cavities through non-collimated, non-relativistic kinetic injections and does not include cosmic rays.}

\subsection{A synergy between weak shocks and X-ray cavities to heat cluster cores}\label{sec:disc_cycle}

Assuming that the TNG-Cluster results are broadly representative of the true AGN-driven shock population in galaxy clusters, and combining these findings with observed shock properties from the literature, {with this paper we are putting forward} a more comprehensive picture of AGN feedback in clusters of galaxies. The prevalence of shocks and X-ray cavities in cool core clusters, their comparable energetics, with observations sometimes showing even stronger shocks, and the observed spatial distribution, where cavities on average lie closer to the SMBH than shocks (seen in both simulations and data, Fig.~\ref{fig:pcav_pshock_rad}), all indicate that no single feedback mechanism solely dominates the heating process. If the AGN outbursts deposit comparable amounts of energy into both shocks and cavity enthalpy, then the dominant mode of feedback likely varies with radius. This points to a complex, scale-dependent feedback landscape in cluster cores, where shocks and cavities work together in synergy to regulate the thermal state of the intracluster gas. 

Although some clusters host only cavities ({or at least appear to do so, 33\%}), we find that half of the shocks occur in clusters that host both (Fig.~\ref{fig:panel}, pie chart). Moreover, shocks and cavities exhibit similar age distributions, on the order of a few tens of Myr (Fig.~\ref{fig:thermo_prop} for shocks, and cavities too, but we do not show it here). This suggests either a common origin from a single outburst or several AGN outbursts occurring on comparable timescales in these clusters.

To better investigate these scenarios, we track the time evolution of these AGN-driven shocks and cavities in one cluster. We use the subbox0, {of the TNG300 simulation \citep[e.g.,][]{TNG_stellar_mass}} with the same physical model as TNG-Cluster, but offering higher temporal resolution (snapshots spaced by <10 Myr). This allows us to capture the immediate gas response to individual SMBH kinetic feedback events for one cluster, the most massive in TNG300 (Halo ID 0, with a mass of $2 \times 10^{15}$ M$_\odot$ at $z=0$). 

Inspecting visually the subbox over time (but staying at low-z $\sim$ 0.2-0.1), we find that, most often, AGN outbursts produce an X-ray cavity and an associated shock front circular pattern. This candidate shock initially surrounds the cavity, but subsequently detaches and propagates ahead of it. Meanwhile, the cavity rises radially away from the SMBH, trailing behind the expanding shock. 

\begin{figure}
    \centering
    \hspace{-0.8cm}
    \includegraphics[width=0.47\textwidth]{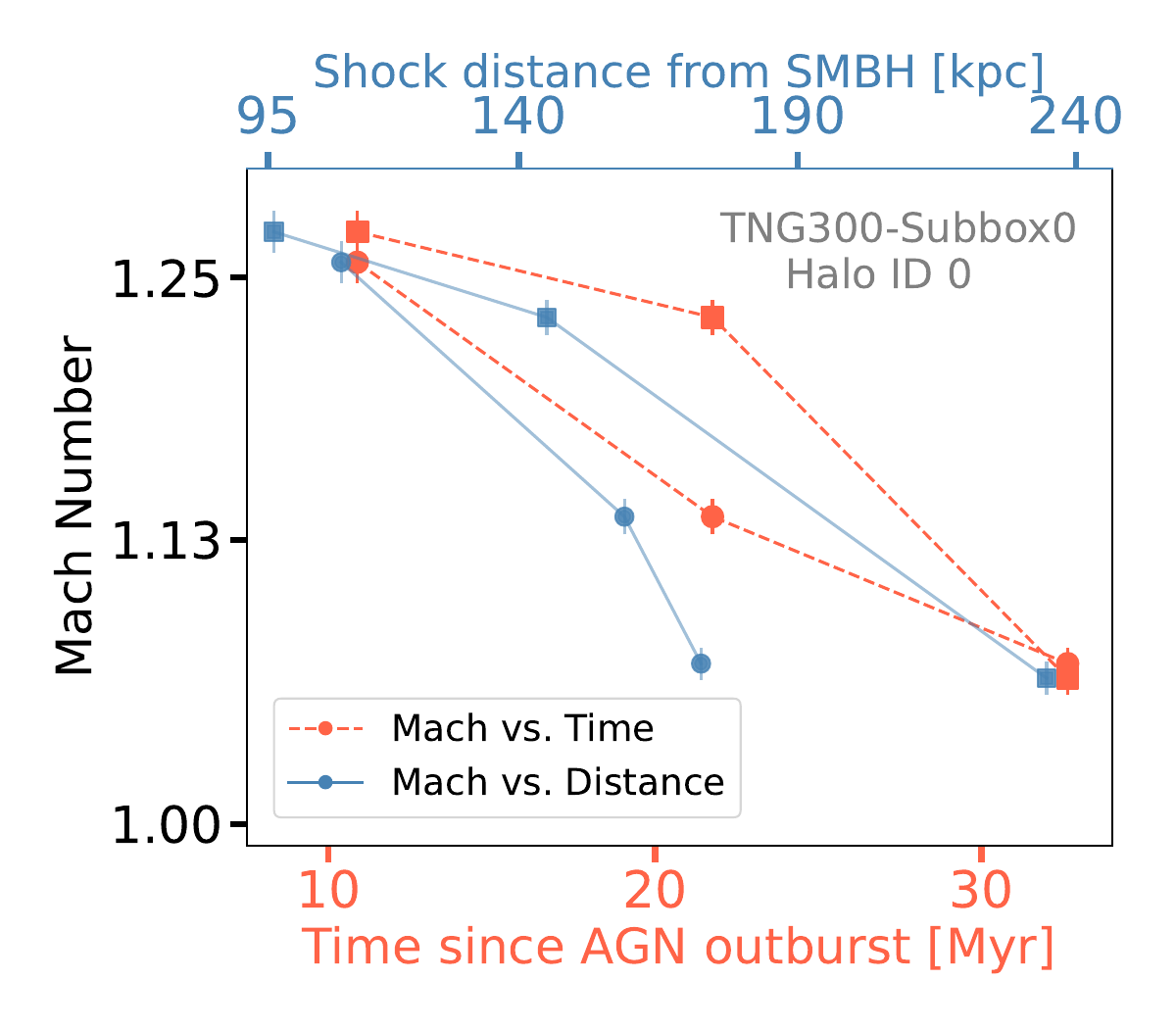}
    \caption{Mach number evolution for two shock events (square and circle markers) in the most massive cluster {from the TNG300 cosmological simulation (subbox0) }halo at low redshift. Initially weak, the Mach number steadily declines with time (orange dotted line) and distance (blue solid line) from the SMBH, approaching 1 before the surface brightness jump vanishes.}
    \label{fig:dicsu_mach}
\end{figure}

Focusing on two representative sequences around $z \sim 0.2$, Fig.~\ref{fig:dicsu_mach} tracks the evolution of distinct shock fronts from their initial detection (as surface brightness jumps in the mock \textit{Chandra}) until they become undetectable. Both features exhibit comparable kinetic powers of $\sim10^{45}$erg s$^{-1}$ during their early evolution. After $\sim$30 Myr, the cavities fade (are not visible in the mock \textit{Chandra} and UM/GGM filtered images), while the shock front remains visible as an arc-like edge in mock \textit{Chandra} images. But, as shown in Fig.~\ref{fig:dicsu_mach}, the Mach number of the shock decreases during the sequence, indicating a transition from a supersonic ($\mathcal{M} \sim$ 1.2) shock to an almost subsonic wave within a few tens of Myr (and across $\sim$ 200 kpc). Although the anticorrelation between Mach number and distance from the central SMBH is weak across the full sample (right panel of Fig.~\ref{fig:panel_shocks}), Fig.~\ref{fig:dicsu_mach} illustrates that, in individual cases, shocks can gradually evolve into sound waves. 

In observations, where X-ray cavities are mostly seen in pairs, a small fraction of clusters host multiple cavity pairs at different radii, tracing successive radio outbursts separated by tens of Myr \citep[e.g., A~262, with four outbursts 28 Myr apart][]{2009Clarke}. Whether a similarly high duty cycle exists for shocks is unclear, as only a few clusters show multiple shocks (e.g., A~2052, RBS~797), while idealized numerical simulations suggest that a single jet episode can produce multiple shock fronts \citep[e.g.,][]{2007Bruggen}.
It is therefore uncertain whether the observed configurations of shocks and cavities arise from the same outburst in observed systems.

Overall, TNG-Cluster shows that X-ray cavities and shocks have offsetted spatial distributions but similar powers. The subbox0 {of TNG300} confirms that most cavities and shocks originate from the same outburst with comparable energy partitioning and rise at different speeds until fading (weak shocks into sound waves). The overall picture points to a powerful, multi-scale heating mechanism where {shocks and cavities }jointly sustain and regulate the thermal balance of the cluster core.

\paragraph*{Sound waves, turbulence, sloshing{, and other core heating channels.}}

\begin{figure}
    \centering
    \includegraphics[width=0.49\textwidth]{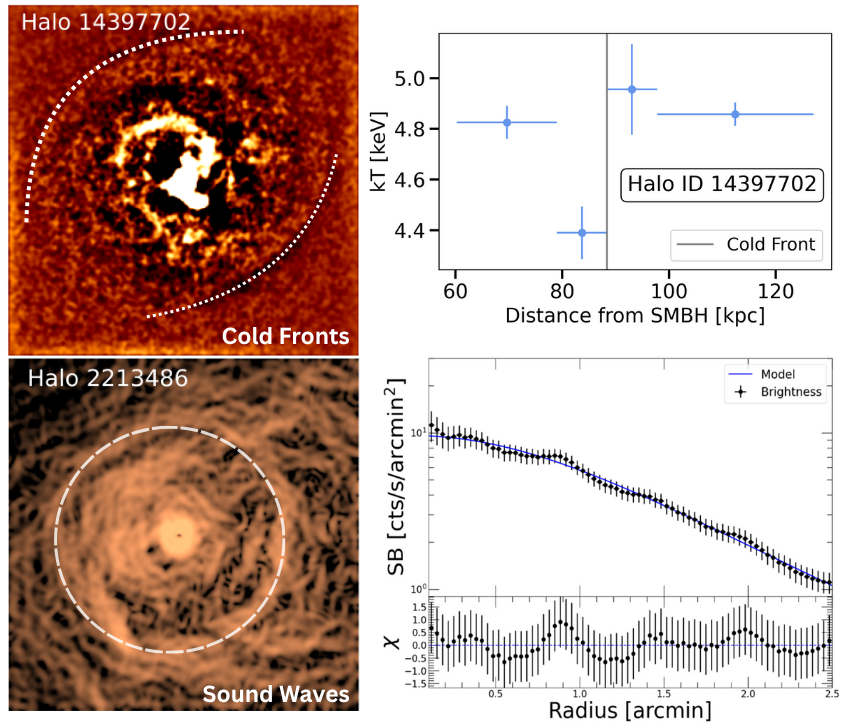}
    \caption{Other typical X-ray structures found in TNG-Cluster: {cold fronts and sound waves}. \textit{Top:} Cold fronts appear as surface brightness edges similar to shocks, but the temperature {gradient is opposite: the gas behind the front is colder, while the gas ahead is hotter.} \textit{Bottom:} Sound waves are characterized by faint concentric ripples, typically only visible in GGM-filtered mock \textit{Chandra} observations and as sinusoidal patterns in the residuals of the surface brightness profiles.}
    \label{fig:dicsu_other_xray_sign}
\end{figure}

Among other mechanisms that may heat the ICM, sound waves generated by AGN activity and/or cavity inflation have been proposed in Perseus and Centaurus \citep[e.g.,][]{Fabian2006_Perseus_Shock,2008Sanders_soundwaves}. These waves appear as 3–10\% pressure enhancements in ripple-like patterns around cluster centers, propagating outward and dissipating their mechanical energy as heat \citep[][]{2017Fabian}. However, even with 1 Ms Perseus data, expected temperature jumps across ripples remain undetected. In TNG-Cluster, we find similar surface brightness features {that look like sound waves.} For example, Halo ID 2213486 shows concentric edges in the GGM-filtered mock \textit{Chandra} image (Fig.~\ref{fig:dicsu_other_xray_sign}), with profiles showing wave-like modulations akin to Perseus. This matches the results of the subbox0 of TNG300 where $\mathcal{M} \sim 1.25$ shocks decay into subsonic waves within $\sim$30 Myr. A detailed analysis of sound wave propagation and associated temperature patterns in TNG-Cluster is left for future work.

Turbulence may contribute to ICM heating \citep{2014Zhuravleva}, but Hitomi and \textit{XRISM} data (e.g., Perseus, Hydra A) show low core velocities ($<165$ km s$^{-1}$), typically insufficient to offset cooling. We defer to future work an analysis of shock- and cavity-driven turbulence in TNG-Cluster, including mock \textit{XRISM} spectroscopy.

Sloshing can redistribute thermal energy by displacing gas via gravitational potential perturbations, reducing central density and X-ray emissivity while increasing cooling time. In the TNG-Cluster mocks, we detect 8 sloshing fronts in 6 of the 100 clusters, some coexisting with shocks. This is likely a lower limit, as many cores {among the simulated clusters} show spiral temperature structures in \cite{2024Lehle}, though we only count features with surface brightness discontinuities in this work. For example, Figure \ref{fig:dicsu_other_xray_sign} (top) shows two cold fronts resembling shocks on the surface brightness map {and} confirmed via spectral analysis. Similar sloshing features are seen in observed systems such as Centaurus A \citep[][]{2016Sanders_CentaurusA_shock} or in A~2052 \citep[][]{2011Blanton_A2052}.

{In addition to the heating mechanisms discussed above, cosmic rays (CRs) released from AGN radio lobes can mix with the dense ICM, and, through various CR energy loss mechanisms, transfer their energy to the thermal gas at a rate that can efficiently heat the gas and suppress cooling flows \citep[e.g.,][]{1969Kulsrud,2008Guo,2017Jacob_Pfommer}. Although CRs are not modeled in TNG-Cluster, they have the potential to play a significant role in regulating cooling flows for real clusters.}

\subsection{Importance of AGN-driven shocks and the need for further studies}

The results presented in this work highlight the need for larger samples and spatially resolved studies of AGN-driven shocks, as they likely contribute significantly to the isotropic heating of cluster cores. 

X-ray cavities, while commonly used to estimate jet power, trace only a portion of the total energy output and may underestimate the true mechanical power. Both observational results \citep[][]{2023Ubertosi_rbs} and our TNG-Cluster analysis suggest that shock-inferred powers correlate with the ICM radiative losses, $L_{\mathrm{cool}}$, consistent with earlier cavity-based studies \citep[e.g.][]{2004Birzan}, but indicate that, on average, $P_{\mathrm{jet,shock}} \gtrsim P_{\mathrm{jet,cav}}$. {Assuming that the observationally-inferred energetics are not biased}, this raises the possibility that cavity-based methods systematically miss a significant fraction of the total AGN energy budget, underscoring the need for more statistics. Moreover, the unknown \textit{Chandra} and \textit{XMM-Newton} selection functions limit current constraints on shock incidence, and a systematic survey is needed to provide an unbiased census of AGN-driven shocks and their role in the feedback cycle.

\section{Summary and conclusions}\label{sec:conclusion}

In this paper, {we have investigated} the prevalence, properties, and impact of AGN-driven shocks in a subsample of 100 galaxy clusters from the TNG-Cluster cosmological simulation. The clusters were mass-selected to match the ten observed clusters with known shocks and available Mach number and power estimates (M$_\text{500c}=1.2$ - $8.5 \times 10^{14}$ M$_\odot$, see Table \ref{tab:shocks}). Using observational analysis techniques, we search{ed} for and characteriz{ed} shock fronts in each simulated cluster using mock \textit{Chandra} ACIS-I observations, generated with a deep 600 ks exposure at redshift $z=0.1$, providing a wide, Mpc-scale field of view. We also leveraged the previously identified population of AGN-driven X-ray cavities in TNG-Cluster halos \citep{Prunier2025a,Prunier2025b} to characterize and quantify the cavity population within this mass-selected subsample.

Our approach uncovered a diverse range of AGN-driven shock morphologies within a cosmological framework and enables direct comparison with the few observed systems hosting shocks. We addressed key open questions regarding the frequency of AGN-driven shocks in clusters, their role in the overall energy budget, and how different feedback signatures vary with cluster radius.
We summarize our main findings in the following points:

\begin{itemize}

    \item Shocks are not rare in the TNG-Cluster mass-selected sample, with about 30 {out of} 100 hosting one or more weak shocks, for a total of 50, all in cool core clusters (Section \ref{subsec:morpho}, Fig.~\ref{fig:panel_shocks}).\\

    \item {The shocks found in the simulated clusters} exhibit a wide range of morphologies, including shocks surrounding expanding X-ray cavities, large-scale cocoons, and arc-like fronts (Fig.~\ref{fig:panel}). {About half of the clusters hosting shocks also contain X-ray cavities, while the remainder show shocks without detectable cavities.} Multiple shocks per cluster are detected, with up to three distinct fronts identified in 15 systems (Section~\ref{subsec:morpho}, Fig.~\ref{fig:panel_shocks}).\\

    \item Detected shocks are weak, with a median Mach number of 1.1, which is consistent, albeit slightly lower, than the median Mach number of 1.2 for the 17 observed shocks (Section \ref{subsec:morpho}, Fig.~\ref{fig:panel}).\\
    
    \item The morphologies, proximity to the central AGN, and prevalence of these shocks in cool core clusters all point to an AGN-driven origin (Section \ref{sec:discussion_represent}). These features result from episodic, high-velocity kinetic energy injections by the central SMBH, similar to the formation of the TNG X-ray cavities \citep{Prunier2025a}.\\

    \item In TNG-Cluster, AGN-driven weak shocks and X-ray cavities carry comparable mechanical power of 10$^{44-46}$ erg s$^{-1}$, while in observations, shocks {have been found to be} typically more powerful than cavities (for 7 out of 10 observed clusters, Fig.~\ref{fig:shock_vs_cav_power}).\\
    
    \item A test using shorter exposure times on mock observations of TNG-Cluster suggests a detection bias in favor of strong $>$10$^{45}$ erg s$^{-1}$ , centrally located shocks (subsection ~\ref{sec:discussion_represent}). Observed shocks are {therefore} likely representing only the high-power tail of the true underlying population.\\

    \item Weak shocks are an efficient heating channel, capable of offsetting gas cooling within the cooling radius (Fig.~\ref{fig:pcav_pshock_lcool}). Unlike cavities, their large arced or cocoon-like morphologies enable them to deliver energy isotropically (Fig.~\ref{fig:panel}). The shocks heat the gas through compression, as evidenced by deprojected increases in temperature and density beneath the shock front  (Fig.~\ref{fig:thermo_prop}), before gradually weakening and transitioning into subsonic sound waves over tens of Myr  (Fig.~\ref{fig:dicsu_mach}).\\

    \item In observations and the TNG-Cluster simulation, the spatial distribution between X-ray cavities and weak shocks shows a radial offset, with cavities typically closer to the central AGN than shocks (Fig.~\ref{fig:pcav_pshock_rad}).
    This suggests that shocks are dominant feedback channels at larger radii, extending from $\sim$ 50 kpc to the cooling radius and beyond. 
\end{itemize}

The overall picture suggested {by TNG-Cluster is that AGN drive }multi-scale heating processes in cluster cores, where shocks and cavities act in concert to regulate the cooling of the intracluster gas. According to the TNG-Cluster predictions, weak shocks are a key component of the AGN feedback loop: they are present in low-entropy, cool core clusters and contribute as significantly as X-ray cavities to the feedback within the cooling region, and beyond. Moreover, these shocks are on average found at larger radii than cavities, making them the dominant feedback mechanism at those distances, where they also operate more isotropically.

The representativeness of shocks in TNG-Cluster relative to the broader, potentially larger population of AGN-driven shocks in real galaxy clusters remains uncertain. Nevertheless, the demonstrated realism of the TNG simulation in reproducing cluster thermodynamic properties, along with its agreement with both large- and small-scale {spatially-resolved} observational signatures (including X-ray cavities), supports the robustness and general applicability of TNG predictions to AGN-driven shocks in clusters.

Although X-ray cavities have long been the primary observational probe for AGN feedback and jet power estimates, our findings suggest they may systematically underestimate the total mechanical energy output. This highlights the critical role of AGN-driven weak shocks and the need for their systematic detection and characterization in large, unbiased cluster samples. The advent of next-generation high-resolution X-ray imaging and spectroscopy missions will be essential to fully test TNG-Cluster predictions, advancing our understanding of AGN feedback and its role in regulating the intracluster gas.

\section*{Acknowledgements}
MP acknowledges funding from the Physics Department of the University of Montreal (UdeM) and the Centre for Research in Astrophysics of Quebec (CRAQ). MP and AP acknowledge funding from the European Union (ERC, COSMIC-KEY, 101087822, PI: Pillepich). FU acknowledges support from the research project PRIN 2022 ``AGN-sCAN: zooming-in on the AGN-galaxy connection since the cosmic noon", contract 2022JZJBHM\_002 -- CUP J53D23001610006. JHL acknowledges funding from the Canada Research Chairs and from the Discovery grant program from the Natural Sciences and Engineering Research Council of Canada (NSERC). 

The TNG-Cluster simulation has been executed on several machines: with compute time awarded under the TNG-Cluster project on the HoreKa supercomputer, funded by the Ministry of Science, Research and the Arts Baden-Württemberg and by the Federal Ministry of Education and Research; the bwForCluster Helix supercomputer, supported by the state of Baden-Württemberg through bwHPC and the German Research Foundation (DFG) through grant INST 35/1597-1 FUGG; the Vera cluster of the Max Planck Institute for Astronomy (MPIA), as well as the Cobra and Raven clusters, all three operated by the Max Planck Computational Data Facility (MPCDF); and the BinAC cluster, supported by the High Performance and Cloud Computing Group at the Zentrum für Datenverarbeitung of the University of Tübingen, the state of Baden-Württemberg through bwHPC and the German Research Foundation (DFG) through grant no INST 37/935-1 FUGG. 

All the analysis and computations associated to this paper have been realized on the Vera cluster of the MPCDF.

\section*{Data Availability}
The TNG simulations, including TNG-Cluster, are publicly available and accessible at \url{www.tng-project.org/data}, as described in \cite{2019NelsonPublicReleaseTNG}. Other data directly related to this publication are available on request from the corresponding author.

\bibliographystyle{mnras}
\bibliography{TNG-Cav-Xray-Comp}

\appendix

\bsp	
\label{lastpage}
\end{document}